\documentclass[apj]{emulateapj}
\usepackage{hyperref}
\usepackage{amsmath}
\usepackage{color}

\shorttitle{Radio Activity in Massive Galaxy Clusters}
\shortauthors{W. Mo et al.}

\begin{document}

\title{The Massive and Distant Clusters of {\it WISE} Survey. VIII. Radio Activity in Massive Galaxy Clusters}

\author{
Wenli Mo\altaffilmark{1},
Anthony Gonzalez\altaffilmark{1},
Mark Brodwin\altaffilmark{2},
Bandon Decker\altaffilmark{2},
Peter Eisenhardt\altaffilmark{3},
Emily Moravec\altaffilmark{1,4},
S. A. Stanford\altaffilmark{5},
Daniel Stern\altaffilmark{3},
and
Dominika Wylezalek\altaffilmark{6}
}

\altaffiltext{1}{Department of Astronomy, University of Florida, Bryant Space Science Center, Gainesville, FL 32611}
\altaffiltext{3}{Jet Propulsion Laboratory, California Institute of Technology, Pasadena, CA 91109}
\altaffiltext{2}{Department of Physics and Astronomy, University of Missouri, 5110 Rockhill Road, Kansas City, MO 64110}
\altaffiltext{4}{Astronomical Institute of the Czech Academy of Sciences, Bo\v{c}n\'{i} II 1401/1A, 14000 Praha 4, Czech Republic}
\altaffiltext{5}{Department of Physics, University of California, Davis, One Shields Avenue, Davis, CA 95616, USA}
\altaffiltext{6}{European Southern Observatory, Karl-Schwarzschildstrasse 2, 85748 Garching, Germany}

\begin{abstract}

We present a study of the central radio activity of galaxy clusters at high redshift. Using a large sample of galaxy clusters at $0.7<z<1.5$ from the Massive and Distant Clusters of {\it WISE} Survey and the Faint Images of the Radio Sky at Twenty-Centimeters $1.4$~GHz catalog, we measure the fraction of clusters containing a radio source within the central $500$~kpc, which we term the cluster radio-active fraction, and the fraction of cluster galaxies within the central $500$~kpc exhibiting radio emission. We find tentative ($2.25\sigma$) evidence that the cluster radio-active fraction increases with cluster richness, while the fraction of cluster galaxies that are radio-luminous ($L_{1.4~\mathrm{GHz}}\geq10^{25}$~W~Hz$^{-1}$) does not correlate with richness at a statistically significant level. Compared to that calculated at $0 < z < 0.6$, the cluster radio-active fraction at $0 < z < 1.5$ increases by a factor of $10$. This fraction is also dependent on the radio luminosity. Clusters at higher redshift are much more likely to host a radio source of luminosity $L_{1.4~\mathrm{GHz}}\gtrsim10^{26}$~W~Hz$^{-1}$ than are lower redshift clusters.
We compare the fraction of radio-luminous cluster galaxies to the fraction measured in a field environment. For $0.7<z<1.5$, we find that both the cluster and field radio-luminous galaxy fraction increases with stellar mass, regardless of environment, though at fixed stellar mass, cluster galaxies are roughly $2$ times more likely to be radio-luminous than field galaxies. 

\end{abstract}


\maketitle

\section{Introduction}

Radio observations are a sensitive probe of the complex physics in the centers of galaxy clusters. Feedback from the mechanical energy of powerful radio jets can regulate star formation in the central region of the galaxy cluster. This scenario has been invoked to counterbalance the lower than expected star formation rate in galaxy clusters with a cooling flow \citep{fabian12}. Many authors have observed radio-mode feedback in action, with large cavities being created by the influx of energy from radio lobes and jets \citep{hlavacek-larrondo13,mcdonald15}.

Several studies have shown that overall Active Galactic Nuclei (AGN) activity is higher in centers of galaxy clusters than in the field \citep[e.g.,][]{galametz09,martini13,bufanda17,mo18}, and is especially pronounced when considering radio-selected AGN. There is evidence that the level of radio activity in clusters is correlated with dynamical state, with dynamically young and merging clusters exhibiting enhanced radio activity \citep{owen99, sobral15, moravec20b}. However, the physical reason for the enhancement of radio AGN activity in galaxy clusters is not well explained. Massive galaxies ($M_{*}>10^{12}~M_{\odot}$) are more likely to be radio-loud \citep[e.g.,][]{best05,seymour07}. Because massive galaxies are also more likely to reside in denser environments, the increase in central radio activity in galaxy clusters could be a reflection of the increase in the average mass of galaxies towards the centers of galaxy clusters \citep[e.g.,][]{joshi17}. Recent galaxy mergers are more likely to trigger radio-loud AGN \citep{chiaberge15}, to which the environments of proto-clusters and galaxy groups are more conducive. \citet{izquierdo-villalba18} find that radio-mode feedback is the main component affecting the stellar build up of the host galaxy, resulting in different galaxy populations in halos with and without radio sources. Thus, understanding the history of radio activity in cluster galaxies may be key to understanding environmental differences in other galaxy properties.

Radio activity in galaxy clusters is often linked to the brightest cluster galaxy (BCG), the most luminous and centrally positioned galaxy in the potential well of the cluster. Its unique location is primed for studies of how galaxy feedback can affect the cluster environment. Studies conducted on the radio activity of BCGs find that they are more likely to be radio-loud than other galaxies of the same stellar mass and than non-BCG cluster galaxies \citep{best07}.
\citet{mittal09} found that BCGs are almost uniformly radio-loud in dynamically relaxed clusters. Furthermore, at least in local clusters, the BCG radio power is correlated with the strength of the cluster's cool core \citep{kale15}.

Radio sources have successfully been used as beacons for finding protoclusters and galaxy clusters. The Clusters Occupied by Bent Radio AGN \citep[COBRA,][]{paterno-mahler17} survey specifically looks for dense environments around bent-tailed radio sources, a tell-tale sign of the interaction between the intracluster medium and radio emission.
\citet{castignani14} search for high-redshift galaxy cluster candidates around Fanaroff-Riley (FR) Type I sources \citep{fanaroffriley}, which are more likely to reside in cluster environments.
The Clusters Around Radio-loud AGN (CARLA) survey exploits exceptionally luminous radio sources to find protocluster candidates at $1.3<z<3.3$ \citep{wylezalek13,wylezalek14,noirot16,noirot18}.
Meanwhile, large catalogs of galaxy clusters found via the Sunyaev-Zeldovich (SZ) flux decrement \citep[e.g.,][]{hasselfield13,bleem15} may be biased by the emission of radio sources still bright in millimeter wavelengths \citep{lin07}.

Most studies of statistical samples of radio sources in galaxy clusters have been conducted at low or intermediate redshift ($z<0.5$). These studies conclude that enhanced radio activity is most common in brightest cluster galaxies, and that the probability of activity depends upon both galaxy stellar mass and cluster richness \citep{best07,sommer11}. Several studies, however, find evidence for evolution in the radio luminosity function, jet power relation, and the cluster radio-loud fraction \citep[e.g.,][]{donoso09,birzan17,lin17}, suggesting that low redshift studies may not fully translate at higher redshift. 

The Massive and Distant Clusters of {\it WISE} Survey (MaDCoWS) is the largest high-redshift galaxy cluster sample, providing a catalog of $2300$ galaxy clusters at $z\sim1$ \citep{gonzalez18}.  \citet[][hereafter M18]{mo18} first studied the cluster AGN population in MaDCoWS galaxy clusters as a function of central cluster distance and cluster richness. M18 found that the surface density of radio-selected AGN is $\sim8$ times higher in the central $1\arcmin$ than in the field, and $\sim4$ times higher than the surface density of optically-selected cluster AGN. 

When considering the fraction of galaxies that are  radio-selected AGN, M18 find that cluster galaxies in the central $1\arcmin$ are $\sim2.5$ times more likely {to exhibit radio activity} than galaxies in the field. The radio-loud fraction is not significantly correlated with the cluster richness. M18 also find that clusters with a radio-selected AGN within $1\arcmin$ of cluster center also showed an enhancement of central optically-selected AGN. For the subset of clusters hosting extended radio sources, \citet{moravec19} and \citet{moravec20} establish a positive correlation between the size of the radio jet and the distance from cluster center, implying a relationship between the density of the intracluster medium (ICM) and confinement of the AGN radio jet. 

In this work, we provide a statistically robust, high-redshift anchor to studies of cluster radio sources by quantifying the prevalence of radio activity in MaDCoWS galaxy clusters at $0.7<z<1.5$, as traced by the most luminous radio AGN.  We focus on two quantities of cluster radio activity: the fraction of clusters with central $500$~kpc radio activity and the fraction of cluster galaxies exhibiting radio emission with $L_{1.4~\mathrm{GHz}}\geq10^{25}$~W~Hz$^{-1}$, which we define as radio-luminous sources. We expand upon the results of M18 to investigate the dependence of cluster radio activity on cluster richness, redshift, radio luminosity, and stellar mass. We also provide a comparison to cluster surveys that specifically target radio sources for discovery. In Section~\ref{sec:data}, we describe the galaxy cluster and cluster radio sources samples. Section~\ref{sec:methods} describes the methods used in this work. The results are presented in Section~\ref{sec:results}. Finally, we discuss the relevance of our results in Section~\ref{sec:discussion}.
Throughout the paper, we adopt the nine-year Wilkinson Microwave Anisotropy Probe (WMAP9) cosmology of $\Omega_\mathrm{M}=0.287$, $\Omega_{\Lambda}=0.713$, and $H_0 = 69.32$~km~s$^{-1}$~Mpc$^{-1}$ \citep{hinshaw13}. Unless otherwise stated, all magnitudes are in the Vega system. The uncertainties are calculated by propagation of Poisson uncertainty. We adopt a canonical spectral index of $\alpha=0.70$ \citep[i.e.,][and references therein]{miley08}.
 
\section{Data}
\label{sec:data}

\subsection{Galaxy Cluster Sample}

The main cluster catalog we refer to is the MaDCoWS cluster survey \citep{gonzalez18}. The MaDCoWS algorithm searches for overdensities in data from the Wide-field Infrared Survey Explorer \citep[{\it WISE},][]{wright10}. MaDCoWS uses a combination of updated AllWISE data \citep{cutri13} in $3.4$~$\mu$m and $4.6$~$\mu$m, with enhanced photometry and astrometry from the All-Sky data release, and optical data from Pan-STARRS Data Release 1 \citep{chambers16} in the Northern Sky or SuperCOSMOS \citep{hambly01} for $\delta<-30^{\circ}$ to reject foreground galaxies. MaDCoWS candidates of highest significance were targeted with {\it Spitzer} IRAC $3.6$ and $4.5$~{$\mu$}m snapshot follow-up. We obtained observations of $200$ MaDCoWS clusters in Cycle 9 and an additional $1759$ in Cycles 11 and 12  (PIs: Gonzalez, Brodwin; PIDs 90177, 11080, 12101).

The additional {\it Spitzer} data enables us to estimate both photometric redshifts and richnesses for the clusters. We summarize the method for these estimations below, and refer the reader to \citet{gonzalez18} for details. Photometric redshifts are determined by comparing the effective {\it Spitzer} $[3.6]-[4.5]$ and Pan-STARRS $i-[3.6]$ colors of cluster galaxies within $1\arcmin$ of cluster center to that expected for a passively evolving galaxy formed at $z=3$. Comparing the photometric redshifts to a sample of $29$ MaDCoWS clusters with spectroscopic redshifts yields a scatter of $\sigma_{z}/(1+z)=0.03$. The scatter does not show strong dependence on redshift \citep[Fig 13,][]{gonzalez18}. The richness is the number of galaxies matching the same color criteria above the $4.5~\mu$m threshold $F_{4.5}>15$~$\mu$Jy ($5\times10^{10}$~M$_{\odot}$) within $1$~Mpc of cluster center. A richness of $\lambda_{15}=22$ corresponds to a cluster mass of $M_{500,c}=10^{14}$~$M_{\odot}$. 

Cluster centroids are determined by using density-smoothed maps of the {\it WISE} overdensity. The cluster center is the pixel containing the peak value after smoothing. The major limitation of this method is the resolution of the smoothed density maps, resulting in a positional uncertainty $\sigma_{\alpha}=\sigma_{\delta}=15\arcsec$. Because the centroid is computed before the photometric redshift is determined, the uncertainty in the photometric redshift does not factor into the uncertainty in the centroid.

Out of $1695$ clusters with {\it Spitzer}-based photometric redshift and cluster richness estimations, $873$ clusters fall within the region covered by FIRST.

\subsection{Radio Data}

The Very Large Array (VLA) carried out the Faint Images of the Radio Sky at $20$~cm (FIRST) survey \citep{becker95}. FIRST offers the largest sky coverage at $1.4$~GHz, complete to $1$~mJy with a resolution of $5\arcsec$. We use the 14 Dec 2017 version of the catalog covering roughly $10,575$ square degrees. The catalog includes radio source position (accurate to $1\arcsec$ at $90\%$ confidence for point sources), integrated flux calculations, and size estimations after Gaussian deconvolution.

In order to avoid edge effects and low quality data in the survey outskirts, we limit the survey region to $110^\circ<\alpha<262^\circ$ and $-8^\circ<\delta<63^\circ$ in the North and $325^\circ<\alpha<40^\circ$ and $-10^\circ<\delta<10^\circ$ in the South. Also, we only consider FIRST sources above the limiting flux ($1$~mJy) and with low probability of being a side lobe ($\mathrm{SIDEPROB}\leq0.015$). After these selection limits, the FIRST catalog is culled to $595,526$ radio sources. We find 213 FIRST sources within 500~kpc of MaDCoWS cluster centers.

\section{Methods}
\label{sec:methods}

\subsection{Cluster Galaxy Catalog}
\label{sec:clugalselect}

Our cluster galaxy selection technique is similar to those developed by \citet{papovich08} and \citet{muzzin13}, further explained in \citet{gonzalez18}. We select for cluster galaxies by using {\it Spitzer} [3.6] and [4.5] data crossmatched with optical photometry from Pan-STARRS. A passively evolving galaxy will become redder in optical-MIR color space between $z=0-2$. We only consider cluster galaxies with $i-[3.6]$ consistent with a passive galaxy within the redshift range $0.7<z<1.5$ for MaDCoWS clusters. We include all galaxies within $500$~kpc of the cluster center with $4.5~\mu$m flux $F_{4.5}>10~\mu$Jy\footnote{$F_{4.5}>10~\mu$Jy corresponds to the 95\% completeness limit in the {\it Spitzer} observations of MaDCoWS clusters.}, $i-[3.6]\geq4.85$, and $5\sigma$ detections in [3.6], [4.5], and Pan-STARRS $i$.

The caveat to this approach is that $\sim40\%$ of galaxies within $500$~kpc and $F_{4.5}>10~\mu$Jy are not detected in Pan-STARRS. We assume these sources to be fainter than the Pan-STARRS limit \citep[$i=23.1$, $5\sigma$][]{chambers16}, which allows us to obtain a lower limit on the $i-[3.6]$ color. Thus, the majority of these sources would also match our selection criteria. We include all galaxies without Pan-STARRS detections but with $F_{4.5}>10~\mu$Jy and S/N$>5$ in [4.5]. Within $500$~kpc of the centers of $873$ clusters within the FIRST footprint, $37,078$ galaxies matched our flux and color criteria.

\subsection{Stellar Mass Estimation}
\label{sec:stellarmassest}

We calculate the stellar masses for cluster galaxies from the {\it Spitzer} $4.5~\mu$m magnitude, adopting the cluster redshift as the redshift of the IR counterpart. We assume a \citet{conroy09} stellar synthesis population (SSP) model and \citet{chabrier03} initial mass function (IMF) simulating a passively evolving galaxy with formation redshift of $z_{f}=3$ \citep{mancone10,wylezalek14}.\footnote{Computed with EzGal  \citep[][\url{http://www.baryons.org/ezgal}]{mancone12}.}

Estimating the stellar mass from [4.5] assumes the flux is stellar-dominated with minimal AGN contribution. To verify that this assumption is reasonable, we examine the distribution of $[3.6]-[4.5]$ colors of the IR counterparts. \citet{stern12} presented an AGN selection based on a MIR color of $[3.6]-[4.5]\geq0.8$. We find only $5$ counterparts ($2.9$\%) have MIR color consistent with the emission being AGN-dominated. We also crossmatch the counterparts to optically-selected quasar catalogs from \citet{richards15}, considering both the master and optical-MIR selected samples from their work. Only $4$ counterparts ($2.3$\%) are identified in the \citet{richards15} quasar catalogs. 

We conclude that AGN emission is not a significant contaminant of the [4.5] stellar mass estimation. To be prudent, we exclude cluster galaxies with $[3.6]-[4.5]>0.6$ from any analysis involving cluster galaxy fractions or stellar mass calculations, i.e. 12\% of counterparts.

\subsection{Cluster Radio Sources Sample Selection}
\label{sec:clusterrs}

In MaDCoWS clusters within the FIRST footprint, we find 213 FIRST radio sources within 500~kpc of cluster centers. To avoid a radio luminosity bias when comparing radio sources across redshift, we impose a luminosity limit on the radio sources. Radio luminosity is calculated as $L_{1.4~\mathrm{GHz}}=4\pi{S_{1.4}}D_{A}^2(1+z)^{3+\alpha}$, where $S_{1.4}$ is the radio flux at $1.4$~GHz, $D_{A}$ is the angular diameter distance, and $\alpha$ is the spectral index, for power law defined as $S_{1.4}\propto\nu^{-\alpha}$. We assume that the radio source is at the cluster redshift.
We only consider radio sources with FIRST luminosity $L_{1.4~\mathrm{GHz}}\geq1\times10^{25}$~W~Hz$^{-1}$, which ensures that a radio source at the FIRST flux limit (1~mJy) can still be detected at the highest cluster redshift under consideration ($z=1.5$). At this luminosity limit, we are most likely only observing radio emission related to the supermassive black hole \citep[e.g.,][]{kellerman16}. Coincidentally, this is also the canonical threshold for FR type I and II sources \citep{fanaroffriley}. Out of the 213 central 500~kpc FIRST sources, 194 were above our luminosity threshold.

We identified {\it Spitzer}$-[4.5]$ counterparts to FIRST radio sources via visual inspection. The visual inspection technique allows us to identify one counterpart for radio sources that were resolved into multiple components in FIRST, decreasing the mis-association of radio sources with complex morphology. We successfully identified 140 IR counterparts for 194 radio sources, where the reduction in the number of identified counterparts to radio sources is due to multiple-component radio sources. We were unable to determine counterparts for 19 radio sources. 

To ensure that the counterpart galaxy is a likely member of the cluster, we apply the same color and magnitude selections as discussed in Section~\ref{sec:clugalselect}. We obtain $i$-band magnitudes for counterparts by crossmatching {\it Spitzer} positions to Pan-STARRS using a $1\arcsec$ radius. A total of 118 counterparts matched the cluster galaxy criteria. 

To estimate the interloper contamination, we use the $i-[3.6]$ color of the radio source counterpart as a proxy for the galaxy's redshift. A passive galaxy at $z\geq0.7$ has $i-[3.6]>4.85$,\footnote{We use Pan-STARRS $i$ magnitudes in AB and $[3.6]$ and $[4.5]$ in Vega, as these are the magnitude systems native to each catalog.} using the same assumptions for stellar synthesis and metallicity as in Section~\ref{sec:stellarmassest}. Thus, counterparts with color below this threshold are most likely at redshifts lower than that of MaDCoWS clusters. We find 12\% of counterparts with color consistent with that of passive galaxies at $z<0.7$ and where emission is not AGN dominated, and therefore are likely foreground interlopers.

\subsection{Cluster Radio Activity Calculations}

The cluster radio active fraction (RAF) is defined as the fraction of galaxy clusters with central radio activity. Any cluster with at least one FIRST radio source above $L_{1.4~\mathrm{GHz}}\geq1\times10^{25}$~W~Hz$^{-1}$ and coincident within the central $500$~kpc cluster center is deemed a ``radio-active galaxy cluster." This luminosity threshold, driven by the depth of the FIRST survey, corresponds to roughly $10\times$ the luminosity of the traditional radio-loudness threshold, and implies that our sources are not just radio loud, but are ``radio luminous". We are only investigating cluster radio activity, and not specifically radio sources, and thus do not require visual identification of the radio source counterpart. We find that $136$ out of $873$ MaDCoWS clusters within the FIRST footprint are radio-active, resulting in a MaDCoWS cluster $\mathrm{RAF}=0.156\pm0.014$, where the uncertainty is calculated from propagation of Poisson error.\footnote{We also perform an alternate estimation of the uncertainty by generating a sets of mock catalogs with the same richness distribution as the real data, randomly assign clusters as being radio-active based upon the active fraction at the cluster richness, and computing the observed RAF. We recover the Poisson uncertainty to two significant figures.}

We define a cluster galaxy as radio-luminous if it matches the criteria for the cluster radio source selection, detailed in Section~\ref{sec:clusterrs}. To summarize, a radio-loud cluster galaxy must match these criteria:
\begin{enumerate}
\item The cluster galaxy matches the color selection described in Section~\ref{sec:clugalselect}.
\item The cluster galaxy is visually identified to be associated with a radio source.
\item The associated radio source has luminosity $L_{1.4~\mathrm{GHz}}\geq1\times10^{25}$~W~Hz$^{-1}$, assuming the photometric redshift of the host galaxy cluster.
\end{enumerate}
The cluster galaxy radio-luminous fraction ($f_\mathrm{rl}$) is then defined as the fraction of radio-luminous cluster galaxies.

In $873$ clusters with a total membership of $37,078$ galaxies, we find $118$ counterparts that matched the cluster galaxy color selection, which we consider to be radio-luminous cluster galaxies. This equates to a total $f_{rl}=(3.18\pm0.29)\times10^{-3}$.

\section{Results}
\label{sec:results}

\subsection{Dependence on Cluster Richness}
\label{sec:clurich}

\begin{figure}
    \centering
    \includegraphics[width=\columnwidth]{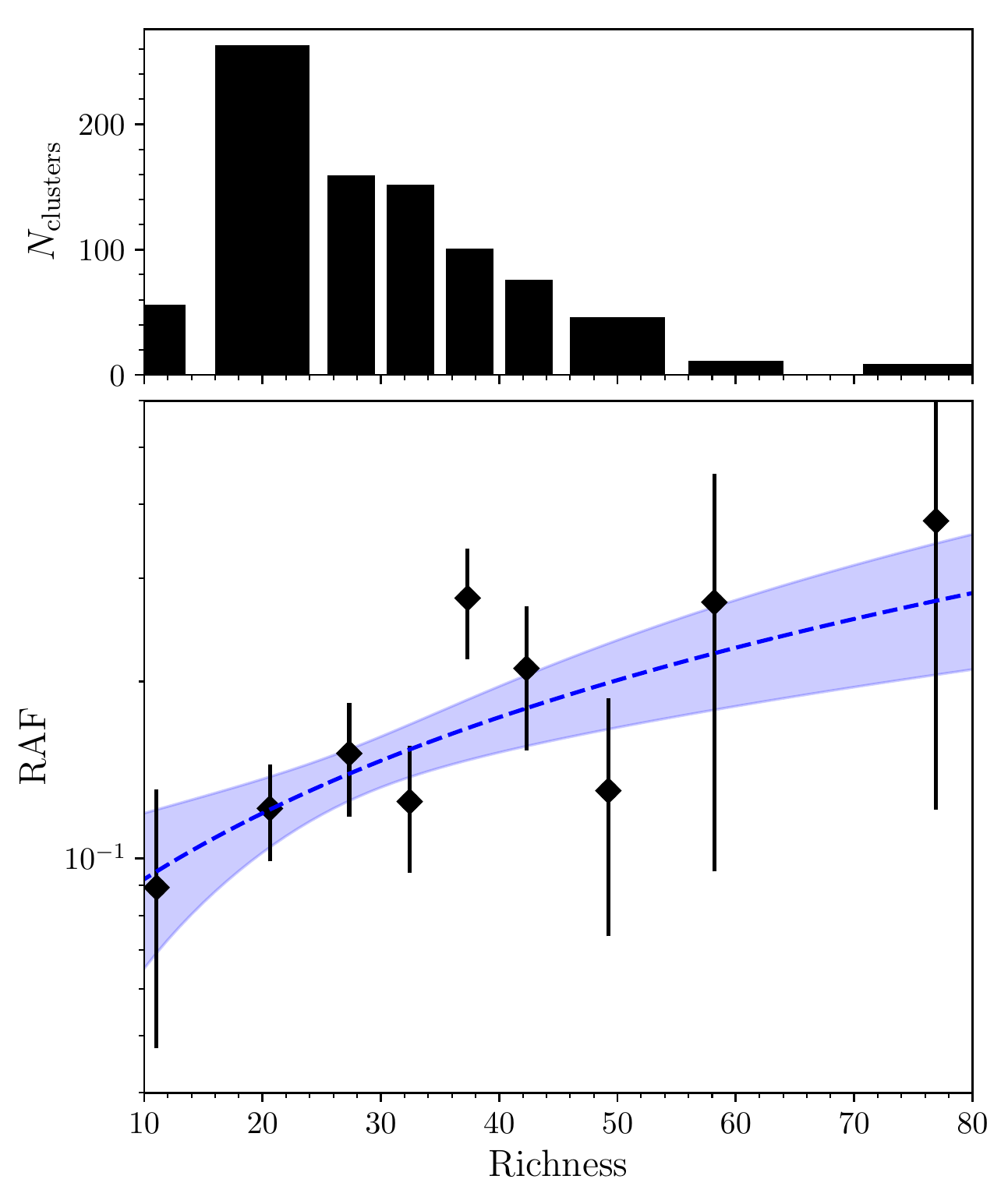}
    \caption{Top: The number of galaxy clusters within each richness bin. Bottom: The cluster RAF as a function of cluster richness. The blue dashed line and shaded region indicates the best-fit linear relationship and uncertainty of that relationship, respectively. The error bars correspond to 1$\sigma$ uncertainties.}
    \label{fig:clusterraf}
\end{figure}

\begin{figure}
    \centering
    \includegraphics[width=\columnwidth]{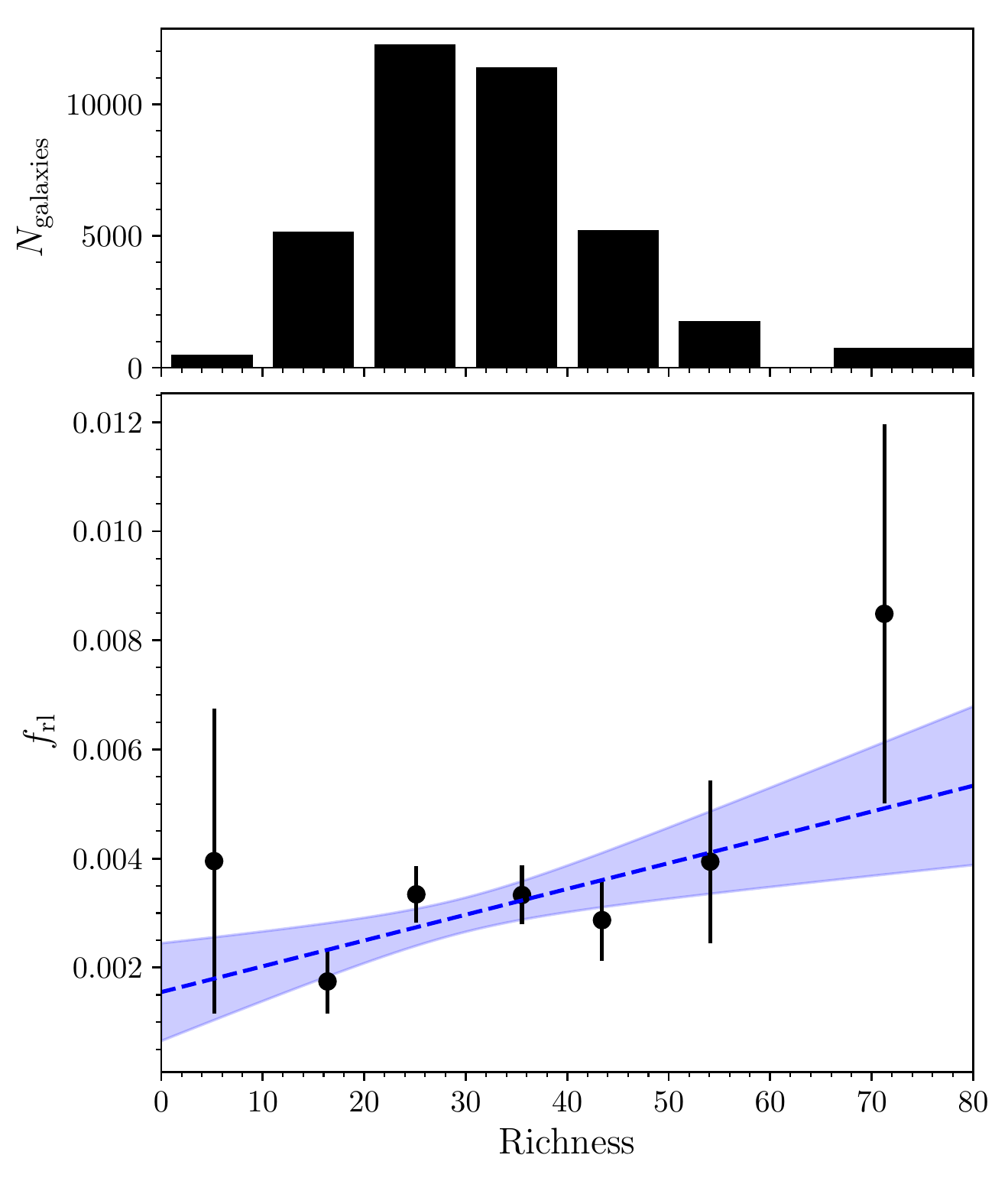}
    \caption{Top: The number of cluster galaxies within the central $500$~kpc of the clusters of each richness bin. Bottom: The cluster galaxy $f_\mathrm{rl}$ as a function of cluster richness. The blue dashed line and shaded region represent the best-fit linear relationship and the uncertainty of that relationship, respectively.}
    \label{fig:frl_richness}
\end{figure}

We investigate the dependence of cluster RAF on the cluster richness and, by proxy, cluster mass. We split the clusters into richness bins spanning $\lambda_\mathrm{15}=0-123$.\footnote{Defined as the number of cluster galaxies above $F_{4.5}>15~\mu$Jy \citep{gonzalez18}.}
We then calculate the cluster RAF in each richness range. 
The number of clusters within each richness bin is shown as the top panel of Figure~\ref{fig:clusterraf}. 

The bottom panel of Figure~\ref{fig:clusterraf} shows the relationship between the cluster RAF and richness. The uncertainties were calculated as the propagation of the Poisson uncertainty of the number of radio-active clusters and the number of total clusters per bin. We fit a linear model of the form $y=ax+b$ to our data via a least-squares method. The best fit results were  $a=(2.7\pm1.2)\times10^{-3}$ and $b=(6.5\pm3.7)\times10^{-2}$, with a reduced $\chi^{2}$=0.92 indicating a reasonable fit.

Our results suggest that the likelihood that a radio-luminous galaxy is within the central $500$~kpc of a galaxy cluster increases with cluster richness. The significance of the trend is marginal (positive slope with $2.25\sigma$ significance). This would imply that more massive galaxy clusters have a higher probability of hosting a radio-luminous galaxy within the central 500 kpc than clusters of lower mass. If confirmed, such a trend could physically be due to the fact that more massive galaxy clusters are, by definition, richer and contain more galaxies. The probability is therefore higher that at least one of the galaxies in the central $500$~kpc is radio-luminous.

In Figure~\ref{fig:frl_richness}, we show the distribution of cluster galaxies within the clusters of each richness bin and the radio-luminous cluster galaxy fraction as a function of cluster richness. The uncertainties on $f_\mathrm{rl}$ were calculated by propagating the Poisson uncertainty on the number of radio-luminous cluster galaxies and the number of total cluster galaxies per richness bin. The calculation of $f_\mathrm{rl}$ accounts for the increase in the number of galaxies for richer clusters. Similar to the cluster RAF, the cluster galaxy $f_\mathrm{rl}$ also shows an increase towards richer clusters. A linear relationship between $f_\mathrm{rl}$ and cluster richness has best-fit parameters of $a=(4.7\pm2.5)\times10^{-5}$ and $b=(1.6\pm0.8)\times10^{-3}$ with reduced $\chi^2=1.0$. While the increase in $f_\mathrm{rl}$ is suggestive that the central cluster environment contributes to the increasing RAF with richness, neither trend is statistically significant.

\subsection{Evolution of the RAF}
\label{sec:clusterrafevolution}

\begin{figure}
    \centering
    \includegraphics[width=\columnwidth]{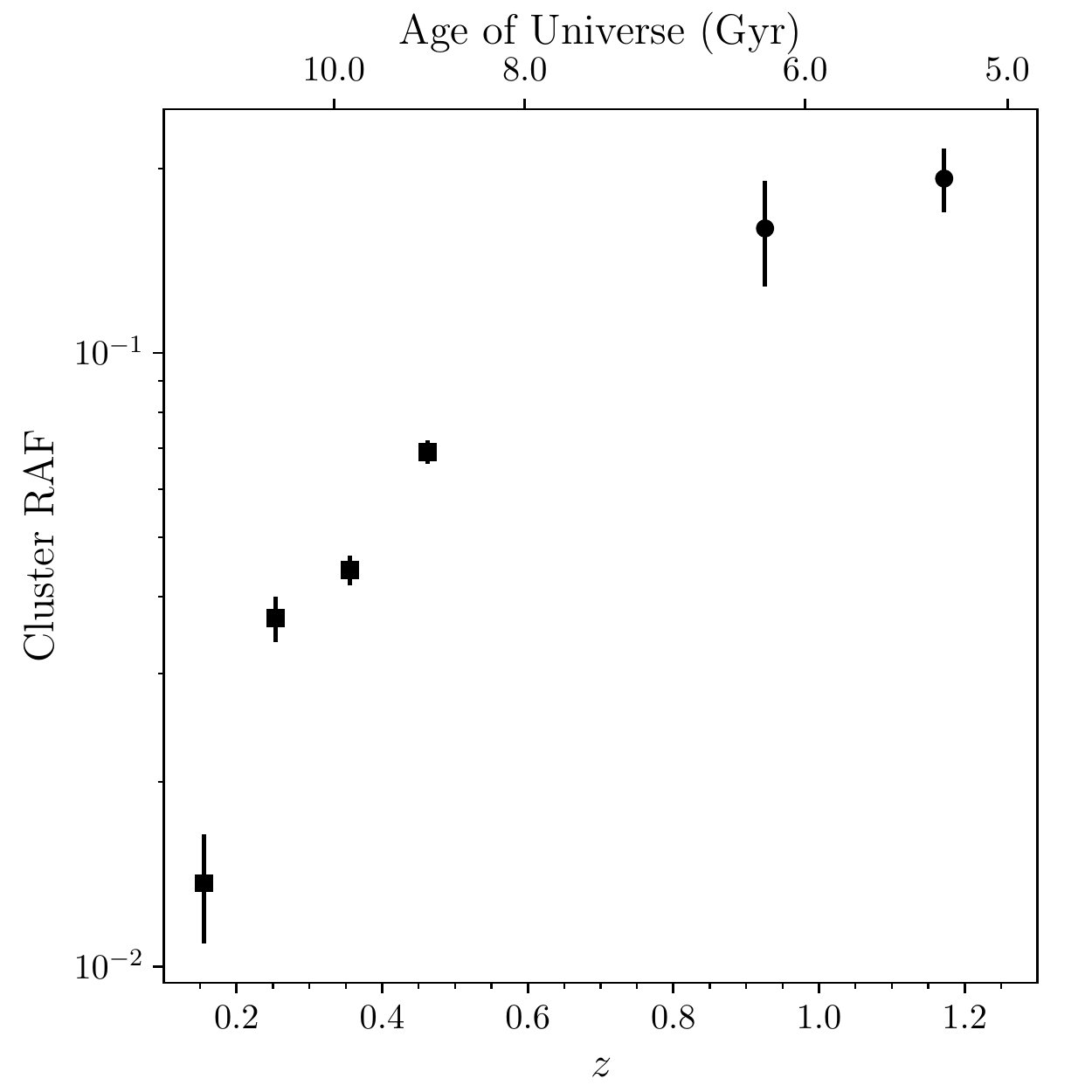}
    \caption{The cluster RAF in redMaPPer (squares) and MaDCoWS (circles) clusters plotted at the mean redshift of each cluster sample. The cluster RAF increases steeply as a function of redshift.}
    \label{fig:clusterrafevolution}
\end{figure}

\begin{deluxetable*}{cccccccc}
\tabletypesize{\footnotesize}
\tablecolumns{8}
\tablecaption{Cluster RAF Redshift Evolution\label{tab:clusterrafevolution}}
\tablehead{\colhead{Cluster} & \colhead{$z$}& \colhead{$\langle{z}\rangle$} & \colhead{$\langle{\lambda}\rangle$} & \colhead{$\langle{M_{500}}\rangle$} & \colhead{Total} &\colhead{Radio Active} & \colhead{Radio} \\ 
\colhead{Catalog} & \colhead{} & \colhead{} & \colhead{} & \colhead{($10^{14} M_{\odot}$)} &\colhead{Clusters} & \colhead{Clusters} & \colhead{Sources}}
\startdata
redMaPPer & $0.08-0.20$ & 0.16 & $33.33\pm0.36$ & $1.72\pm0.02$ & 1830  & 25 & 32 \\
 & $0.20-0.30$ & 0.25 & $32.83\pm0.26$ & $1.69\pm0.01$ & 3874 & 143 & 191 \\
 & $0.30-0.40$ & 0.36 & $33.06\pm0.18$ & $1.7\pm0.01$ & 7520 & 333 & 440  \\
 & $0.40-0.60$ & 0.46 & $46.71\pm0.21$ &  $2.44\pm0.01$ & 8334  & 575 &744 \\
MaDCoWS & $0.77-1.00$ & 0.93 & $33.5\pm0.67$ & $2.35\pm0.1$ & 262 & 37 & 52  \\
 & $1.00-1.50$ & 1.17 &  $33.96\pm0.43$ & $2.4\pm0.06$ & 594  & 95 & 121 
\enddata
\tablecomments{Column 2: Redshift bin. Column 3: Mean cluster redshift within bin. Column 4: Mean cluster richness within bin. Column 5: Mean cluster mass within bin. Column 6: Number of galaxy clusters within redshift bin. Column 7: Number of clusters with central $500$~kpc radio activity. Column 8: Number of FIRST radio sources within $500$~kpc of radio active clusters.}
\end{deluxetable*}

We investigate how the cluster RAF has evolved with cosmic time by comparing MaDCoWS to lower redshift clusters. The redMaPPer cluster algorithm finds galaxy clusters via their red sequence \citep{rykoff14}. We use the redMaPPer catalog for SDSS Data Release 8, which contains $25,000$ galaxy clusters ranging in redshift of $0<z<0.6$. The catalog includes cluster photometric redshift and richness estimate. The redMaPPer richness estimate is based on the methods developed in \citet{rozo09} for maxBCG galaxy clusters. Each galaxy within some cutoff radius of the cluster $R_{c}$ is given a probability of cluster membership based on optically-observed properties like color and magnitude. The cluster richness is defined as the sum of the probability of membership for all galaxies within $R_{c}$ above a luminosity threshold \citep[also see][for details]{rykoff12,rykoff14}. To compare redMaPPer and MaDCoWS clusters, we use the mass-richness relations for both samples (Fig. 11 and Eq. B6 in \citet{rykoff12} and Fig. 16 and Eq. 2 in \citet{gonzalez18}) to define a threshold corresponding to $M_{500,c}\gtrsim1\times10^{14}~M_{\odot}$. For redMaPPer, this corresponds to richness $\lambda_\mathrm{redMaPPer}>20$ and for MaDCoWS, a richness of  $\lambda_{15}>22$.
We split the redMaPPer catalog into 4 redshift bins at $z=0.2$, $z=0.3$, and $z=0.4$ and MaDCoWS into 2 bins at $z=1.0$. The number of total and radio-active clusters per redshift interval is listed in Table~\ref{tab:clusterrafevolution}.

To calculate the RAF in redMaPPer clusters, we again identify all FIRST radio sources with $L_{1.4~\mathrm{GHz}}>10^{25}$~W~ Hz$^{-1}$ that lie within $500$~kpc of the cluster center. As with the MaDCoWS clusters, we define a cluster as being radio-active if there is a FIRST source satisfying these criteria.

The results of the redshift evolution of the cluster RAF are shown in Figure~\ref{fig:clusterrafevolution}. We find that the RAF rises as a function of redshift, increasing more than tenfold between $0.2<z<1.2$. Our results indicate that radio-luminous galaxies in the centers of clusters are much more abundant in the earlier universe than in the present.

The increase in RAF towards higher redshift may in part be due to the inherent differences in cluster populations at different redshifts, for example in their recent star formation or cluster merger history. Though we control for cluster mass when comparing redMaPPer and MaDCoWS clusters, this does not constitute an evolutionary sequence due to cluster mass growth. However, our aim is to compare equivalent mass clusters at different epochs.

We investigate the effects of Malmquist bias at the highest redshift bins (where the radio flux limit is faintest) to quantify how much we may be overestimating radio source counts. We simulate the flux for a sample of sources well below the $1$~mJy flux limit, using a power-law distribution constructed from the FIRST catalog and assigning a flux uncertainty. We then calculate the ratio of the number of simulated sources to the number of observed sources above each flux limit equivalent to luminosity $L_{1.4~\mathrm{GHz}}=1\times10^{25}$~W~Hz$^{-1}$ at each redshift bin. If we assign an uncertainty from a Gaussian distribution centered around $0.5$~mJy with standard deviation $0.1$~mJy, we find that the ratio at $z=1.0$ and $z=1.5$ is $1.1$ and $1.5$, respectively. The increase in the cluster RAF from $z=0.5-1.0$ and $z=0.5-1.2$ is on the factor of $\sim2-2.5$. Thus, even though the effects of Malmquist bias could be driving up the RAF at higher redshifts, the increase towards higher redshifts would still exist even taking into account Malmquist bias.

\subsection{Dependence on Radio Luminosity}
\label{sec:rafradiolum}

\begin{figure*}
    \centering
    \includegraphics[width=\textwidth]{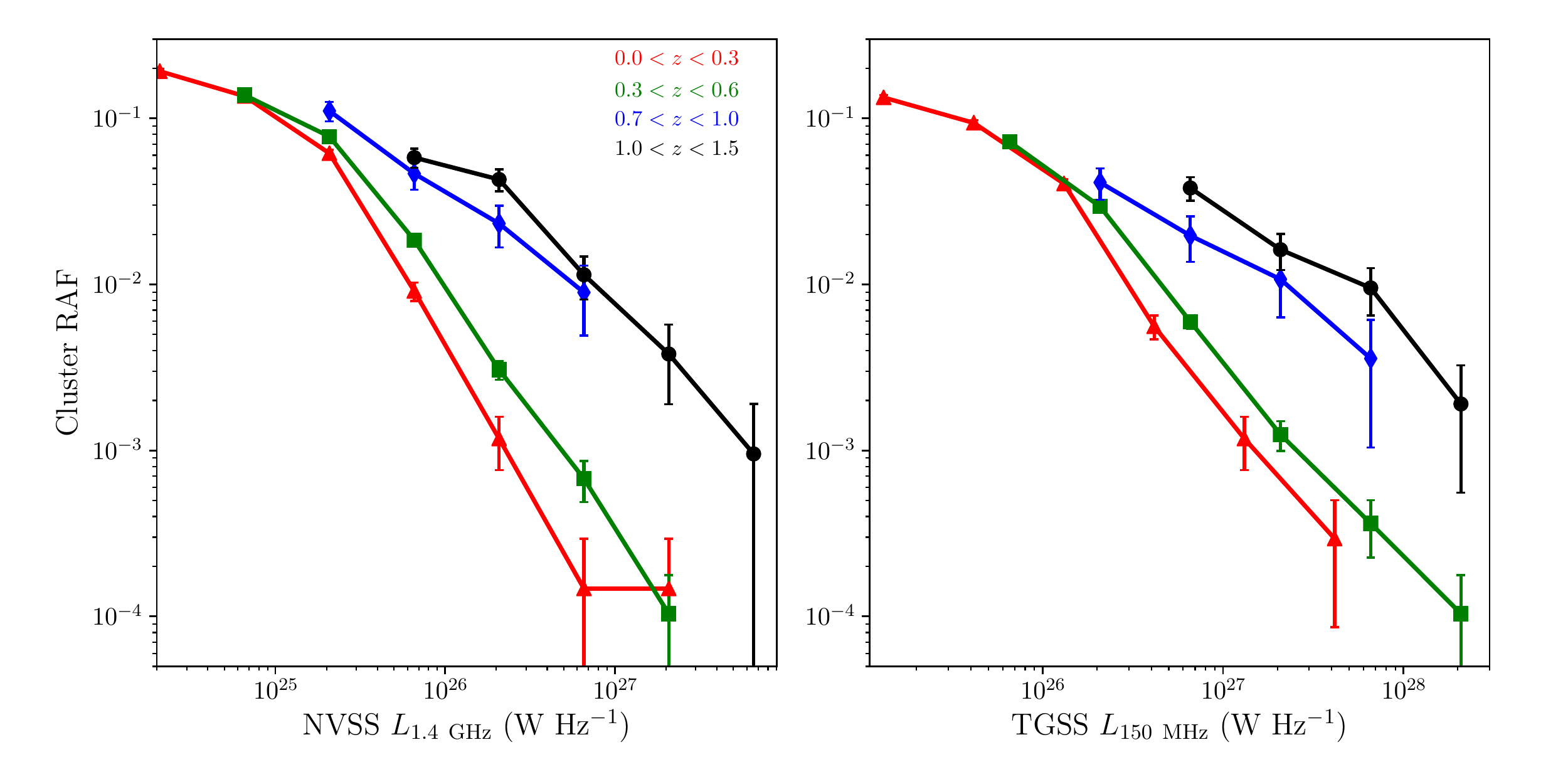}
    \caption{Left: The cluster RAF as a function of $1.4$~GHz radio luminosity of the central radio source. Clusters are binned by redshift. Right: Same as left panel but for $150$~MHz radio luminosity.}
    \label{fig:raf_lum}
\end{figure*}

\begin{figure}
    \centering
    \includegraphics[width=\columnwidth]{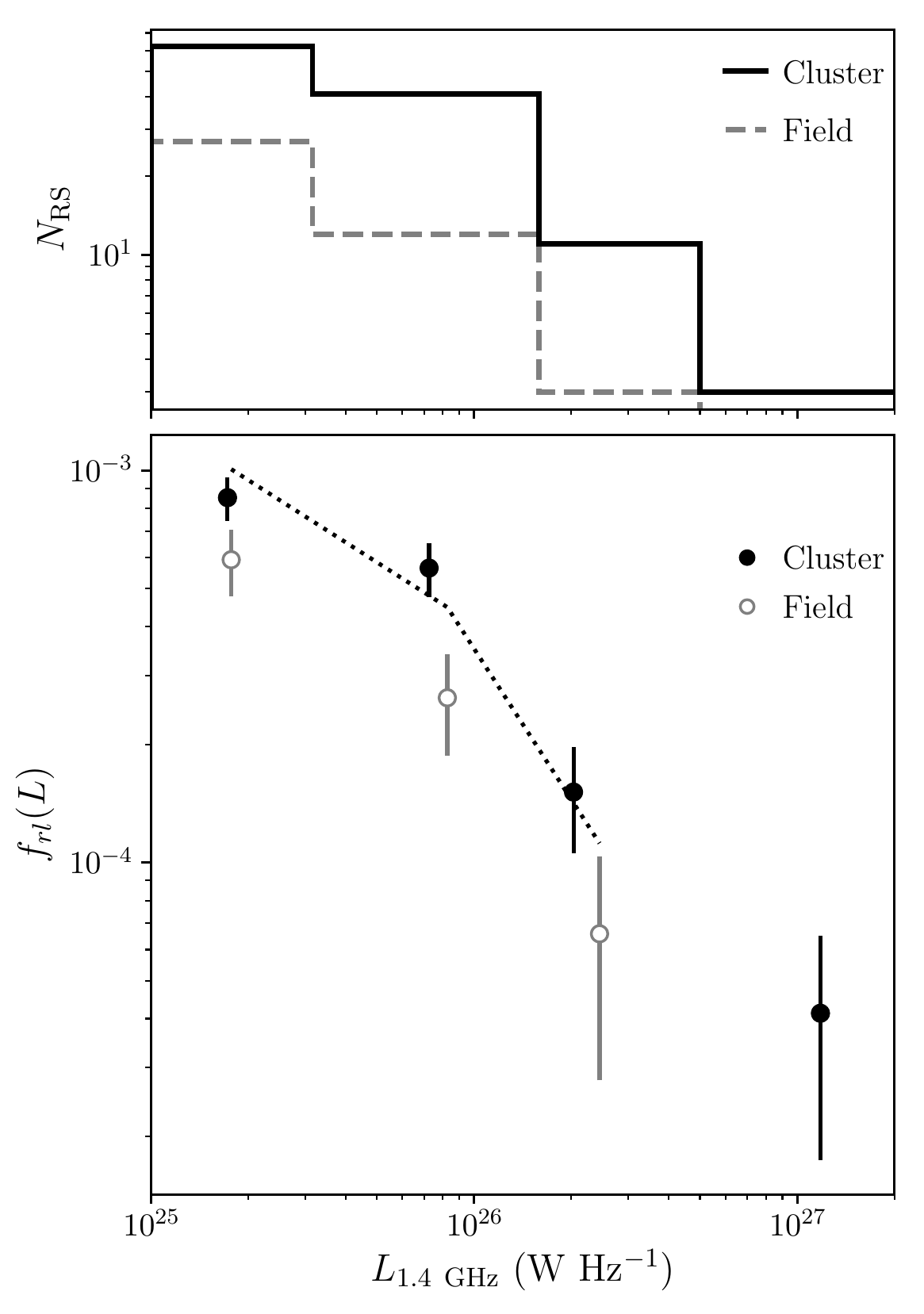}
    \caption{Top: The number of radio sources in the cluster (black solid) and the field (gray dashed) for each luminosity bin. Bottom: RLF for FIRST radio sources in the cluster (black filled) compared to in the SDWFS field (gray open). The dotted line represents the field $f_\mathrm{rl}$ values scaled by $1.70\pm0.31$, the average ratio between the field and cluster $f_\mathrm{rl}$.}
    \label{fig:rlf}
\end{figure}

We investigate the dependence of the cluster RAF upon radio luminosity. We calculate the cluster RAF as a function of radio luminosity, RAF($L$), where we consider radio sources located in the central 500~kpc with luminosity within a set luminosity bin. Instead of using FIRST luminosity, we opt for data from the NRAO VLA Sky Survey \citep[NVSS,][]{condon98}, which surveyed the radio sky north of $\delta>-40^\circ$ at $1.4$~GHz. Though at lower resolution than FIRST, NVSS has a FWHM of $45^{\arcsec}$, allowing for more accurate flux estimation of extended sources that may be resolved into multiple components in FIRST, and larger sky coverage. 

One disadvantage of NVSS compared to FIRST is the higher flux limit. NVSS is only complete to $2.5$~mJy, and we adjust our luminosity limit accordingly. Another is the increased chance of confusion, especially in clusters with multiple radio sources. When calculating RAF, we only consider if the cluster has at least one radio source in the central $0.5$~Mpc. We expect confusion to be minimal in lower redshift clusters due to the higher flux limit. Of the 136 radio-active MaDCoWS clusters identified with FIRST, 92\% of those clusters were identified as radio-active using NVSS, and all NVSS-identified radio-active clusters were also identified as radio-active in FIRST.
Therefore, we do not expect radio source confusion to affect the calculation of the RAF.

Our results from Section~\ref{sec:clusterrafevolution} motivate us to also consider the redshift while investigating the cluster RAF as a function of radio luminosity. We again draw upon the redMaPPer clusters as a low redshift comparison. We split the redMaPPer clusters into redshift bins of $0<z\leq0.3$ and $0.3<z\leq0.6$ and MaDCoWS into bins of $0.7<z\leq1.0$ and $1.0<z\leq1.5$. There are $6802$, $19257$, $559$, and $1050$ clusters within the footprint of NVSS in the lowest to highest redshift bins, again only considering clusters with richness equivalent to $M_{500}>1\times10^{14}~M_{\odot}$.

The left side of Figure~\ref{fig:raf_lum} shows the cluster RAF for NVSS $1.4$~GHz luminosity in bins of $0.5$ dex, considering the clusters of different redshift. The luminosity sensitivity changes as a function of redshift, so we only calculate the cluster RAF to the luminosity given the flux limit $2.5$~mJy and highest redshift cluster within the bin. The minimum luminosity considered is $L_{1.4~\mathrm{GHz}} = 10^{24}$, $10^{24.5}$, $10^{25}$, and $10^{25.5}$~W~Hz$^{-1}$ from lowest to highest redshift bins. We only plot the cluster RAF if at least one cluster contains an NVSS radio source within the luminosity bin. Figure~\ref{fig:raf_lum} shows that the cluster RAF is dependent on the redshift, where higher redshift clusters have a higher fraction of clusters hosting a central radio source of $L_{1.4~\mathrm{GHz}}>3\times10^{25}$~W~Hz$^{-1}$. 

To add a low frequency comparison, we also calculate the cluster RAF as a function of radio luminosity at $150$~MHz with data from the TIRF Giant Metrewave Radio Telescope (GMRT) Sky Survey (TGSS) Alternative Data Release 1 \citep[ADR1,][]{intema17}. TGSS is currently the highest resolution low-frequency survey, covering the entire sky above $\delta=-50^{\circ}$ with FWHM $25^{\arcsec}\times25^{\arcsec}$ and $25^{\arcsec}\times25^{\arcsec}/\cos19^{\circ}$ north and south of $\delta>-19^{\circ}$, respectively. The survey threshold is $25$~mJy.\footnote{$S_{150~\mathrm{MHz}}=25$~mJy is equivalent to $S_{1.4~\mathrm{GHz}}=5$~mJy assuming $\alpha_{150-1400}=0.7$.} We are thus sensitive to sources above $\log{L_{150~\mathrm{MHz}}}=24.8$, $25.5$, $26$, and $26.5$ from lowest to highest cluster richness bin. The number of clusters per redshift bin remains the same as for NVSS. The right panel of Figure~\ref{fig:raf_lum} shows the cluster RAF as a function of $150$~MHz luminosity for clusters of $0<z<1.5$. The cluster RAF as a function of 150~MHz radio luminosity shows a similar trend to that at 1.4~GHz. 

We next construct a radio luminosity function (RLF) at $1.4$~GHz for radio sources in MaDCoWS clusters by calculating the fraction of cluster galaxies as a function of radio luminosity. 
We include a field RLF comparison. To calculate the $f_\mathrm{rl}$ in the field, we use the {\it Spitzer} Deep, Wide-Field Survey (SDWFS, Ashby et al. 2009) with galaxy photometric redshifts  from \citet{chung14}. We select SDWFS galaxies of the same flux ($[4.5]>10~\mu$Jy, $M_{*}>3\times10^{10}~M_{\odot}$) and color criteria as cluster galaxy candidates, as described in Section~\ref{sec:clugalselect}. We also limit the sample to galaxies with photometric redshifts of $0.7<z<1.5$ to match the photometric redshift range of MaDCoWS clusters. There were $3$ MaDCoWS galaxy clusters that fall within the SDWFS field of view. We exclude the $49$ SDWFS galaxies within $1\arcmin$ of these galaxy clusters. A total of $45,520$ SDWFS galaxies matched these criteria. Crossmatching the SDWFS galaxies with FIRST using a $2\arcsec$ crossmatching radius, we find 61 radio-luminous galaxies in SDWFS, where the crossmatched FIRST luminosity is $L_{1.4~\mathrm{GHz}}\geq1\times10^{25}$~W~Hz$^{-1}$, calculated assuming the photometric redshift of the galaxy.  The redshift distribution in cluster galaxies and field galaxies is well matched, as expected from using the cluster galaxy color selection criteria. The mean redshift of radio-luminous MaDCoWS cluster galaxies is $z=1.09$ while that of the radio-luminous field galaxies is $z=1.15$.

The cluster and field RLFs are plotted in Figure~\ref{fig:rlf}. 
The top panel of Figure~\ref{fig:rlf} shows the number of cluster and field radio sources per luminosity bin. The fraction of radio galaxies as a function of FIRST 1.4~GHz luminosity is shown in the bottom panel for both clusters and the field. In the field, the most luminous galaxy had luminosity $L_{1.4~\mathrm{GHz}}=4\times10^{26}$~W~Hz$^{-1}$. Thus, we can only calculate field $f_\mathrm{rl}$ up to 
$L_{1.4~\mathrm{GHz}}<10^{26.5}$~W~Hz$^{-1}$. 

The radio-luminous fractions for both field and cluster galaxies decrease as a function of increasing radio luminosity. However, the radio-luminous fractions in cluster galaxies are higher than that in field galaxies at all radio luminosities. We also show the field radio-luminous fraction scaled by the ratio of the average radio-luminous fraction between cluster and field (dotted line). The average radio-luminous fraction is simply the radio-luminous fraction calculated for all sources in the luminosity range of the field radio sources ($3\times10^{25}{\leq}L_{1.4~\mathrm{GHz}}<5.0\times10^{26}$~W~Hz$^{-1}$). The ratio between average cluster and field radio-luminous fractions is $1.70\pm0.31$. We calculate an R-squared score of $0.84$, indicating that scaling the field by a constant is a reasonable representation of the radio activity in the cluster environment.

\subsection{Dependence on Stellar Mass}
\label{sec:clufracstellarmass}

\begin{figure}
    \centering
    \includegraphics[width=\columnwidth]{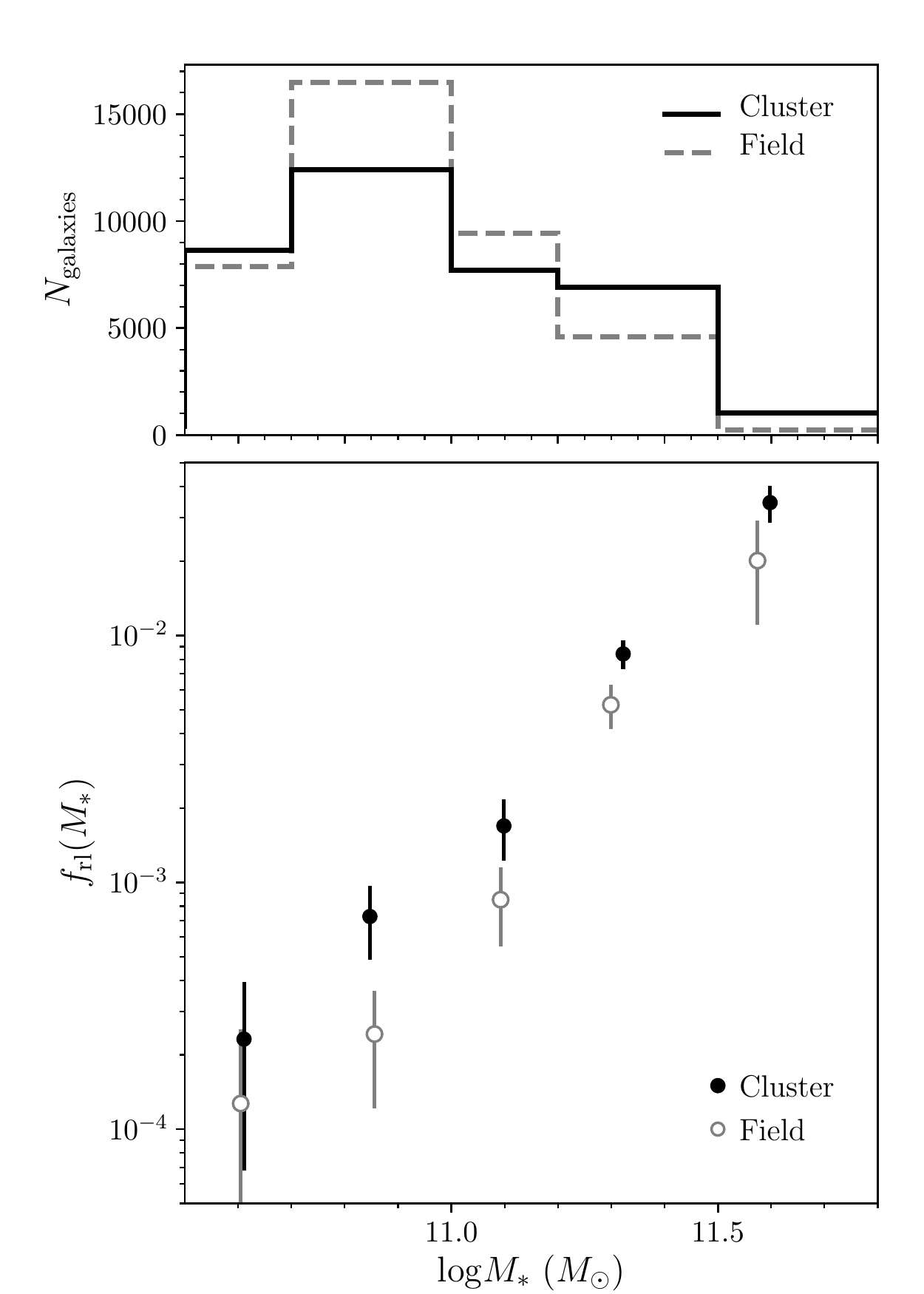}
    \caption{Top: The number of cluster galaxies (black solid) and galaxies matching the same criteria in the field (gray dashed) per stellar mass bin. Bottom: The cluster galaxy $f_\mathrm{rl}$ (black filled) as a function of stellar mass of the central radio source counterpart. The cluster galaxy RAF increases with increasing stellar mass. The field galaxy $f_\mathrm{rl}$ (gray open) also increases with stellar mass, but remains lower than that in the cluster. The average ratio between cluster and field $f_\mathrm{rl}$ is $3.6\pm0.6$.}
    \label{fig:raf_stellarmass}
\end{figure}

We next try to quantify the effects of the environment on triggering radio activity. 
To identify radio-luminous galaxies in SDWFS, we apply the cluster galaxy color and flux criteria (described in Section~\ref{sec:clugalselect}) and crossmatch the selected galaxies to FIRST radio source positions with a $2\arcsec$ crossmatching radius and apply the same $L_{1.4~\mathrm{GHz}}>10^{25}$~W~Hz$^{-1}$ selection, calculated assuming the galaxy's photometric redshift and spectral index of $\alpha=0.70$. 
For $0.7<z<1.5$, we find $42$ radio-luminous galaxies in SDWFS, or an overall radio-luminous fraction in the field of $(9.22\pm0.14)\times10^{-4}$.

The bottom panel of Figure~\ref{fig:raf_stellarmass} shows $f_\mathrm{rl}$ for galaxies in MaDCoWS clusters and in SDWFS as a function of stellar mass. The number of field and cluster galaxies per bin are shown in the top panel. The $f_\mathrm{rl}$ increases with increasing stellar mass by a factor of $\sim150$ in clusters and field between $10^{10.5}<M_{*}\lesssim10^{11.7}$~$M_{\odot}$. The average ratio between cluster and field $f_\mathrm{rl}$ is $3.6\pm0.6$ for $10^{10.5}<M_{*}\lesssim10^{13}$~$M_{\odot}$.

Our results are in agreement with those of M18, who find that the radio-luminous\footnote{The radio-luminous fraction was referred to as the radio-loud fraction in M18, though the definitions are the same.} fraction in the innermost $0.5$~Mpc of MaDCoWS clusters is $\sim3$ times higher than that in the field. They also find the field-relative radio-luminous fraction in $10.5<\log(M_{*})\leq11.6$ and $\log(M_{*})>11.6$ galaxies are comparable, albeit slightly higher in the former $M_{*}$ range.

\section{Discussion}
\label{sec:discussion}

Overall, our analysis has shown that radio activity is not ubiquitous in the centers of galaxy clusters, and is dependent on the cluster mass, redshift, and galaxy stellar mass.

\subsection{Effects of Cluster Richness}

In Section~\ref{sec:clurich}, we find tentative evidence that cluster RAF appears to increase as a function of cluster richness, though we cannot statistically constrain that the $f_\mathrm{rl}$ also increases with richness. M18 find an increase in the central radio-luminous fraction (presented as the radio-selected AGN fraction) as a function of increasing cluster richness, though the radio-luminous fractions of each richness bin are comparable to each other within the central $1\arcmin$. In this work, we benefit from visually-identified radio source counterparts instead of a matching radius, physical as opposed to angular distances, and a uniform luminosity cut rather than a flux limit. Despite these differences, our results remain consistent with those of M18.

Many studies have previously found a link between cluster richness and radio activity. Both observational \citep[e.g.,][]{hatch14} and theoretical \citep[e.g.,][]{orsi16} studies have shown that radio sources trace denser environments than mass-matched counterparts without evidence for radio activity. This relationship extends to the fraction of BCGs with radio activity. Though we do not identify BCGs in our clusters, there is a high likelihood that much of the observed radio activity is attributed to the BCG given that we only consider the central $500$~kpc (though see Sections~\ref{sec:clusterrafevolution} and \ref{sec:disc_clusterrafevolution}). 

Multiple studies find a similar mass dependence in lower redshift clusters. In $z\leq0.2$ clusters, \citet{lin07} find that the radio active fraction of BCGs in $\log{M_{200} (M_{\odot})}>14.2$ clusters are higher than that in lower mass clusters, for $\log{L_{1.4~\mathrm{GHz}}}>24$ and BCGs brighter than $M_{K}=-24$. Similarly, in $z<0.3$ clusters, \citet{stott12} find that the fraction of clusters with radio-luminous BCGs is $0.10\pm0.04$ in clusters with X-ray temperature $T_{X}<2.4$~keV, equivalent to $M_{500}<1.4\times10^{14}$~$M_{\odot}$. This fraction rises to $0.38\pm0.09$ in clusters of higher X-ray temperatures, considering radio luminosities to $L_{1.4~\mathrm{GHz}}=2\times10^{23}$~W~Hz$^{-1}$. \citet{bird08} also cite a correlation between increasing richness in local galaxy groups and probability of extended radio emission in BCGs.

While the above studies do see such a trend, one exception is \citet{best07}. Studying the radio-loud BCG fraction in 625 nearby optically-selected groups and clusters, they do not observe a strong dependence of the fraction of radio-loud BCGs and the cluster's velocity dispersion. Also, the redshift dependence is not previously well-constrained. \citet{gralla11} find that the radio-loud BCG fraction increases more significantly for clusters between $0.35<z<0.65$ than in their sample at $0.65<z<0.95$, implying that the radio active fraction as a function of richness could be dependent on redshift.

\subsection{Cluster RAF Evolution}
\label{sec:disc_clusterrafevolution}

We show in Section~\ref{sec:clusterrafevolution} and Figure~\ref{fig:clusterrafevolution} that the cluster RAF increases by more than tenfold between $0.2<z<1.2$. Many authors have noted an increase in the frequency of luminous radio sources towards higher redshifts in both cluster and non-cluster environments \citep[e.g.,][]{lin07,sommer11,yuan16}.  \citet{donoso09} construct the radio luminosity function (RLF) of the MegaZ luminous red galaxy catalog between $0.4<z<0.8$ and compare with RLF of local galaxies at redshift range $0.03<z<0.3$ \citep{best05,best05a,best06}. They find that the number of radio sources with $L_{1.4~\mathrm{GHz}}>10^{24}$~W~Hz$^{-1}$ increases by a factor of 1.5 between the lowest and highest redshifts while the number of higher luminosity radio sources with $L_{1.4~\mathrm{GHz}}>10^{26}$~W~Hz$^{-1}$ increases tenfold. For radio sources in clusters in the redshift range $0.3<z<1.2$, \citet{birzan17} find that the fraction of clusters that host a cluster radio source with $L_{843~\mathrm{MHz}}\geq10^{26}$~W~Hz$^{-1}$ is $\sim7$ times higher in $z>0.6$ than in $z<0.6$. Our results agree with previous studies, and add that the frequency of cluster radio sources with luminosity $L_{1.4~\mathrm{GHz}}>10^{25}$~W~Hz$^{-1}$ continues to increase out to $z=1.2$.

It is interesting to compare these results for evolution in the cluster RAF with the evolution in the fraction of radio-luminous BCGs. \citet{gralla11} found that, for a luminosity threshold of $L_{1.4~\mathrm{GHz}}=4.1\times10^{24}$~W~Hz$^{-1}$, there is negligible evolution in the fraction of radio-luminous BCGs out to $z=0.95$. Similarly, \citet{lin17} find little evolution in the radio-luminous BCG fraction at $0.3<z<1.2$ for $L_{1.4~\mathrm{GHz}}=5.0\times10^{24}$~W~Hz$^{-1}$.
At the same luminosity threshold as \citet{lin17}, we find that the cluster RAF increases by a factor of $3$ from $z=0.2$ to $z=1$. 

The above results are not contradictory, but rather indicative that the observed evolution is driven by an increase in radio activity among the non-BCG galaxy population with increasing redshift. Indeed, \citet{gralla11} found a $2.9\sigma$ increase in the radio-loud fraction of non-BCG cluster galaxies at $0.65<z<0.95$ compared to $0.35<z<0.65$, which translates into a higher cluster RAF at high redshift. However, studies have different radius and luminosity limits, which may also contribute to the different results between studies.

\subsection{Dependence on Radio Luminosity}

The increase in the number of luminous radio sources towards higher redshift clusters, as stated in Section~\ref{sec:rafradiolum} and shown in Figure~\ref{fig:raf_lum}, is most likely due to the shift in the RLF as a function of redshift. Many studies have cited an evolution in the RLF towards higher radio luminosities in radio-selected sources in the field. Studying a sample of VLA-COSMOS radio-selected AGN, \citet{smolcic09} find that the RLF for low-luminosity ($L_{1.4~\mathrm{GHz}}\sim10^{25}$~W~Hz$^{-1}$) radio AGN is $3$ times higher at $0.9<z\leq1.3$ than at $0.1<z\leq0.35$. \citet{donoso09} find a factor of $2$ increase for radio AGN samples at $\langle{z}\rangle=0.55$ compared with $\langle{z}\rangle=0.14$ for $L_{1.4~\mathrm{GHz}}=10^{25}$~W~Hz$^{-1}$, and this increases to a factor of $8$ for $L_{1.4~\mathrm{GHz}}=10^{26}$~W~Hz$^{-1}$. 

Extending to higher redshifts and luminosities, \citet{rigby11}, using steep-spectrum ($\alpha>1.06$) radio sources only, find that the RLF increases by a factor of $2.5$ for $\log(L_{1.4~\mathrm{GHz}})=25.5$ and a factor of $15$ for $\log(L_{1.4~\mathrm{GHz}})=26.3$ between $z=0.2-1.0$ (see their Figure 9). In our results, the cluster RAF at $z=1.2$ is a factor of $2$ higher than that at $z=0.2$ at $\log(L_{1.4~\mathrm{GHz}})=25.5$ and $20$ for $\log(L_{1.4~\mathrm{GHz}})=26.3$, which is on par with previous results found for the field RLF. However, our results highlight that more luminous radio sources are more likely to reside in higher redshift galaxy clusters, and that higher redshift clusters must experience associated radio-mode feedback.

\subsection{Dependence on Stellar Mass}

In Section~\ref{sec:clufracstellarmass} and Figure~\ref{fig:raf_stellarmass}, we find that both higher galaxy mass and the cluster environment drive increased cluster radio activity. Comparing the environments around radio galaxies and radio-loud quasars at $1.3<z<3.2$ to the environments around radio-quiet galaxies matched in mass and redshift, \citet{hatch14} found that the radio-loud population traces denser environments than do the radio-quiet, and that the cluster environment enhances the likelihood of radio jet formation in galaxies. Our results quantify these findings that the cluster environment is more conducive to radio activity in galaxies than the field environment by an overall factor of $3.6$.

It is also worth noting that the enhancement of the field-relative radio-luminous fraction in clusters is not limited to the highest stellar mass cluster galaxies. The $f_\mathrm{rl}$ is higher in clusters versus field over the entire stellar mass range in consideration. 
For $M_{*}<10^{11.2}~M_{\odot}$, the ratio of the cluster to field level is $2.6\pm0.8$.
This indicates that all galaxies in the cluster have a higher probability for being radio-luminous, not just the BCG. Similar BCG-only studies are most likely missing the environmental component of radio activity enhancement.

\subsection{Duty Cycle of Radio Activity in Galaxy Clusters}

The cluster RAF can be directly applied to the duty cycle of radio emission in galaxy clusters, defined to be the fraction of time that a galaxy cluster has at least one radio source with $L_{1.4~\mathrm{GHz}}>10^{25}$~W~Hz$^{-1}$ in the central $500$~kpc region. We find that the cluster radio emission duty cycle is $15$\% for radio luminosity $L_{1.4~\mathrm{GHz}}>10^{25}$~W~Hz$^{-1}$, assuming that all clusters alternate on-off states \citep[though see][on radio AGN lifetimes]{hardcastle19}. However, we showed that the cluster RAF is dependent on cluster richness and redshift. Richer clusters would experience a longer time frame or greater fraction of time where the radio emission is deposited onto the central cluster environment. Nearby clusters are less likely to host powerful radio sources in the central regions compared to $z\sim1$ clusters, though the cluster RAF is also shown to depend on radio luminosity. 

\subsection{Extrapolation to Radio-Selected Cluster Surveys}

Radio sources have been used as beacons for galaxy cluster surveys \citep[e.g.,][]{castignani14, wylezalek13, paterno-mahler17}. Because the MaDCoWS sample is unbiased towards clusters with radio sources, we can constrain the rarity of galaxy clusters selected by their central radio source. Specifically, we apply the selection functions of the CARLA and COBRA surveys, which target radio sources with high luminosity and bent radio morphology, respectively.

CARLA obtained {\it Spitzer} snapshot imaging of the environments around 421 luminous radio-loud AGN at $1.2<z<3.2$ \citep{wylezalek13,wylezalek14}. They found that 55\% of radio-loud AGN were in overdense regions, and likely to trace high redshift cluster/protocluster environments. The radio-loud targets considered by CARLA had 500~MHz luminosity $L_{500~\mathrm{MHz}}\geq10^{27.5}$~W~Hz$^{-1}$. We convert NVSS $1.4$~GHz luminosity to 500~MHz luminosity assuming $\alpha_{150-1400}=0.70$, the average spectral index for radio sources in clusters. We consider NVSS instead of FIRST in this case because of the larger coverage area of NVSS. Also, since we are interested in the most luminous radio sources, the higher resolution and lower flux limit of FIRST do not benefit our analysis. We only find 3 galaxy clusters ($0.2$\%) with a radio source above $L_{500~\mathrm{MHz}}\geq10^{27.5}$~W~Hz$^{-1}$. We do not correct for the fact that CARLA environments are most likely less massive and more distant than MaDCoWS clusters, factors that both impact the cluster RAF, although in opposite senses. To first order though, we can infer that the clusters discovered by CARLA represent a small faction of the total cluster population in a uniformly-selected cluster sample. 

The COBRA survey \citep{paterno-mahler17} searched for dense environments around FIRST radio sources with tails bent by the cluster ICM. Bent-tailed radio sources were visually identified in FIRST contours, avoiding the most obvious low-redshift sources and verifying the status of the bent-tail. \citet{wing11} successfully identify $653$ bent-tailed sources. We adopt the same visual inspection approach to identifying bent-tailed radio sources. However, because we consider higher redshift radio sources, FIRST data will be less resolved than that considered in \citet{wing11}. We find 6 MaDCoWS clusters containing a possible bent-tail radio source. This implies that only $0.4$\% of MaDCoWS clusters would have been discovered by targeting bent-tailed sources. Thus both CARLA and COBRA detect a very specific subpopulation of clusters and protoclusters.

\subsection{Contamination of SZ Surveys}

\begin{deluxetable}{cccc}
\tabletypesize{\footnotesize}
\tablecolumns{4}
\tablecaption{Cluster SZ Contamination \label{tab:sze_contamination}}
\tablehead{\colhead{$\nu$ (GHz)} & \colhead{$\alpha=0.5$}& \colhead{$\alpha=0.7$} & \colhead{$\alpha=1.0$}}
\startdata
\multicolumn{4}{c}{}\\
\multicolumn{4}{c}{20\% Contamination}\\ \hline\\
90 & 63 (7.2\%) & 33 (3.8\%) & 9 (1.0\%) \\
150 & 48 (5.5\%) & 22 (2.5\%) & 6 (0.7\%) \\ \hline
\multicolumn{4}{c}{}\\
\multicolumn{4}{c}{50\% Contamination}\\ \hline\\
90 & 30 (3.4\%) & 14 (1.6\%) & 5 (0.6\%) \\
150 & 23 (2.6\%) & 9 (1.0\%) & 3 (0.3\%) \\ \hline
\multicolumn{4}{c}{}\\
\multicolumn{4}{c}{100\% Contamination}\\ \hline\\
90 & 15 (1.7\%) & 8 (0.9\%) & 3 (0.3\%) \\
150 & 9 (1.0\%) & 5 (0.6\%) & 1 (0.1\%)
\enddata
\tablecomments{Total number and percentage of MaDCoWS clusters where the converted $1.4$~GHz flux of the central $1$~Mpc radio source is equivalent to 20\%, 50\%, and 100\% of the expected negative cluster SZ signal at $90$ and $150$~GHz.}
\end{deluxetable}

The flux of a radio source can fill in the SZ decrement from a galaxy cluster. We have established that radio sources are more likely to be in the centers of galaxy clusters. Thus, there is the potential for radio sources to affect and possibly even overwhelm the signal of their host galaxy cluster.

We use the method described in \citet{gupta17} to estimate a galaxy cluster's SZ decrement as radio fluxes at the 90 and 150 GHz bands commonly used by SZ surveys. We also employ the SZ-mass relation of \citet{arnaud10}.  Together, these predict the (negative) flux from the galaxy cluster. We then convert the $1.4$ GHz flux from the central radio source to the flux expected at SZ frequencies. For clusters that contain more than one central radio source, we use the combined flux from those sources. 

The spectral index $\alpha$ chosen to convert from $1.4$ GHz to SZ frequencies is the largest uncertainty in this calculation. \citet{lin07} find that the spectral index for $1.4-4.85$~GHz for non-BCG radio sources is $0.47\pm0.15$. Improving upon the methods developed in \citet{lin07}, \citet{lin09} calculate a steeper mean $1.4-4.85$~GHz spectral index of $0.754\pm0.024$ for $139$ galaxies within $r_{200}$ of the centers of $z<0.25$ galaxy clusters. At higher frequencies, they find that 60\% of sources flatten above $8$~GHz and a third of sources remain steep from $4.9-43$~GHz.
Further, \citet{baek16} report a difference in spectral index as a function of the cluster's dynamical state. In a sample of $10$ cluster AGN in clusters at $z\sim0.02-0.10$ observed at $4.85$~GHz and $22$~GHz , they observed that cluster AGN in non-cool-core clusters had steeper spectral indices ($\alpha\gtrsim1.0$) than those in relaxed clusters ($\alpha\lesssim1.0$). 
For the radio population not limited to those within galaxy clusters, \citet{gralla14} find that the spectral index for $1.4-4.8$~GHz and $1.4-218$~GHz steepens with increasing average flux of the radio source, from $\alpha\sim0.5$ at $S_{1.4}=10$~mJy to $\alpha\sim0.9$ at $S_{1.4}=90$~mJy. Given the uncertainty in the spectral index towards higher frequencies, we compute the calculation using a range of assumed spectral indices. We choose spectral indices of $\alpha=[0.5,0.7,1.0]$ to represent the range of spectral indices reported from $1.4$~GHz to higher frequencies.

The total number and percentage of MaDCoWS clusters that contain central $1$~Mpc radio sources with radio flux above 20\%, 50\%, and 100\% of the cluster's expected SZ decrement is listed in Table~\ref{tab:sze_contamination}. We list these values as a function of the assumed $1.4-90$~GHz and $1.4-150$~GHz spectral index.  At $90$ GHz ($150$ GHz), no more than $1.7$\% ($1.0$\%) of MaDCoWS clusters contain central $1$~Mpc radio sources that overwhelm the expected signal from SZ.

We compare our results to that of \citet{lin09}, who find the distribution of spectral indices between multiple frequencies for 139 radio sources to determine the RLF, then generate central $r_{200}$ radio sources in dark matter halos using a Monte Carlo method to extrapolate a contamination percentage at $145$~GHz. They estimate $\sim0.1\%$ of $M_{200}=10^{14}~M_{\odot}$ clusters at $z=1.1$ would be $100\%$ contaminated by the flux of the central radio source and $0.5\%$ of clusters at the $20\%$ contamination level. For $M_{200}=10^{15}~M_{\odot}$ clusters, the percentages are at negligible amounts for both contamination levels. Comparing to our results at $150$~GHz, we find a much higher fraction of contaminated clusters, at both the $20\%$ and $100\%$ contamination level. This is likely due to the discrepancy in the evolution of radio galaxies in clusters relative to the evolution assumed in \citet{lin09}. They assume that the cluster radio source density increases by a factor of $2$ between $z=0.25-1$, while in this work, we find that the increase is a factor of $4$-$5$. We also assume a straight spectral index between $1.4$~GHz and $90$~GHz and $150$~GHz, when there could be steepening and flattening of the flux between those frequencies. Nevertheless, we observationally demonstrate that the number of MaDCoWS clusters that would have been missed by SZ surveys is larger than previously estimated, but still small.

Our analysis indicates that SZ-based cluster masses are expected to be biased low by at least $20$\% for $\sim1-7\%$ of MaDCoWS-like clusters, and about half of these will be biased low by 50\% or more.  Beyond suffering from biased mass estimates, samples selected in the SZ may have higher incompleteness near the survey S/N limit due to partial fill-in of the decrement.  Cosmological measurements using cluster abundances should account for both incompleness and mass bias due to emission from radio-luminous AGN in high redshift clusters.

\section{Summary}
\label{sec:summary}

We investigated the occurrence of radio activity in the central $500$~kpc region of $1695$ massive galaxy clusters at $z\sim1$. Our main results are as follows:
\begin{itemize}
\item The MaDCoWS radio active fraction (RAF), defined as the fraction of clusters with a radio source within the central $500$~kpc and $L_{1.4~\mathrm{GHz}}\geq1\times10^{25}$~W~Hz$^{-1}$, is $\mathrm{RAF}=0.156\pm0.014$. The MaDCoWS radio-luminous fraction, defined as the fraction of cluster galaxies within $500$~kpc with radio luminosity  $L_{1.4~\mathrm{GHz}}\geq1\times10^{25}$~W~Hz$^{-1}$, is $f_\mathrm{rl}=(3.18\pm0.29)\times10^{-3}$.
\item We find marginal ($2.25\sigma$) evidence that cluster RAF increases with cluster richness. More massive clusters are more likely to contain radio-luminous galaxies in the central region.
\item The cluster RAF evolves strongly with redshift, from $z=0.16$ to $z=1.17$. The cluster RAF is higher at higher redshifts, for all luminosity thresholds, implying that the most luminous radio sources are more likely to reside in distant clusters.
\item More distant galaxy clusters are more likely to host high luminosity radio sources. The fraction of clusters between $1.0<z<1.5$ with a central 500~kpc radio source of $L_{1.4~\mathrm{GHz}}>10^{26.5}$~W~Hz$^{-1}$ is $\sim100$ times higher than that in $0<z<0.3$ clusters.
\item The probability for a galaxy to be  radio-luminous depends upon both stellar mass and environment. Though the fraction of radio-luminous galaxies, $f_\mathrm{rl}$, increases by a factor of $\sim150$ between $\log{M_{*}}=10.5-11.7$ in both cluster and field galaxies, the cluster galaxy $f_\mathrm{rl}$ is $1.5-2.9$ times higher than mass-matched galaxies in the field.
\item We estimate that no more than $1.7$\% and $1.0$\% of MaDCoWS clusters contain a central radio source that could fully overwhelm the expected SZ decrement at $90$~GHz and $150$~GHz, respectively. This fraction is larger than previous estimates, but remains small.
\item SZ-based cluster masses are expected to be biased low due to the partial fill-in of the SZ decrement by radio-luminous AGN. We estimate that the cluster masses for $\sim1-7\%$ ($\sim1-3\%$) of MaDCoWS-like clusters would be biased low by 20\% (50\%).
\end{itemize}

Our findings point toward a scenario where increased radio activity in clusters coincides with the epoch of cluster assembly, during which the effects of radio emission on galaxy evolution within galaxy clusters would be enhanced. The effects of enhanced radio feedback during this time period highlight the importance of understanding the relationship between radio activity and the properties of their host galaxies and the clusters in which they reside.

\section*{Acknowlegements}

Funding for this work is provided by the NASA Astrophysical Data Analysis Program, award NNX12AE15G, the National Science Foundation grant AST-1715181, and through NASA grants associated with {\it Spitzer} observations (PID 90177 and PID 11080).

This publication makes use of data products from the Wide-field Infrared Survey Explorer, which is a joint project of the University of California, Los Angeles, and the Jet Propulsion Laboratory/California Institute of Technology, funded by the National Aeronautics and Space Administration. This work is based in part on observations and archival data obtained by the Spitzer Space Telescope, which is operated by the Jet Propulsion Laboratory, California Institute of Technology under a contract with NASA. Support for this work was provided by NASA through an award issued by JPL/Caltech. The Pan-STARRS1 Surveys \citep[PS1,][]{chambers16} and the PS1 public science archive have been made possible through contributions by the Institute for Astronomy, the University of Hawaii, the Pan-STARRS Project Office, the Max-Planck Society and its participating institutes, the Max Planck Institute for Astronomy, Heidelberg and the Max Planck Institute for Extraterrestrial Physics, Garching, The Johns Hopkins University, Durham University, the University of Edinburgh, the Queen's University Belfast, the Harvard-Smithsonian Center for Astrophysics, the Las Cumbres Observatory Global Telescope Network Incorporated, the National Central University of Taiwan, the Space Telescope Science Institute, the National Aeronautics and Space Administration under Grant No. NNX08AR22G issued through the Planetary Science Division of the NASA Science Mission Directorate, the National Science Foundation Grant No. AST-1238877, the University of Maryland, Eotvos Lorand University (ELTE), the Los Alamos National Laboratory, and the Gordon and Betty Moore Foundation.


\begin{thebibliography}{}
\expandafter\ifx\csname natexlab\endcsname\relax\def\natexlab#1{#1}\fi

\bibitem[{{Arnaud} {et~al.}(2010){Arnaud}, {Pratt}, {Piffaretti},
  {B{\"o}hringer}, {Croston}, \& {Pointecouteau}}]{arnaud10}
{Arnaud}, M., {Pratt}, G.~W., {Piffaretti}, R., {et~al.} 2010, \aap, 517, A92

\bibitem[{{Baek} {et~al.}(2016){Baek}, {Chung}, {Tremou}, {Sohn}, {Jung}, \&
  {Ro}}]{baek16}
{Baek}, J., {Chung}, A., {Tremou}, E., {et~al.} 2016, Astronomische
  Nachrichten, 337, 82

\bibitem[{{Becker} {et~al.}(1995){Becker}, {White}, \& {Helfand}}]{becker95}
{Becker}, R.~H., {White}, R.~L., \& {Helfand}, D.~J. 1995, \apj, 450, 559

\bibitem[{{Best} {et~al.}(2006){Best}, {Kaiser}, {Heckman}, \&
  {Kauffmann}}]{best06}
{Best}, P.~N., {Kaiser}, C.~R., {Heckman}, T.~M., \& {Kauffmann}, G. 2006,
  \mnras, 368, L67

\bibitem[{{Best} {et~al.}(2005{\natexlab{a}}){Best}, {Kauffmann}, {Heckman},
  {Brinchmann}, {Charlot}, {Ivezi{\'c}}, \& {White}}]{best05}
{Best}, P.~N., {Kauffmann}, G., {Heckman}, T.~M., {et~al.} 2005{\natexlab{a}},
  \mnras, 362, 25

\bibitem[{{Best} {et~al.}(2005{\natexlab{b}}){Best}, {Kauffmann}, {Heckman}, \&
  {Ivezi{\'c}}}]{best05a}
{Best}, P.~N., {Kauffmann}, G., {Heckman}, T.~M., \& {Ivezi{\'c}}, {\v Z}.
  2005{\natexlab{b}}, \mnras, 362, 9

\bibitem[{{Best} {et~al.}(2007){Best}, {von der Linden}, {Kauffmann},
  {Heckman}, \& {Kaiser}}]{best07}
{Best}, P.~N., {von der Linden}, A., {Kauffmann}, G., {Heckman}, T.~M., \&
  {Kaiser}, C.~R. 2007, \mnras, 379, 894

\bibitem[{{Bird} {et~al.}(2008){Bird}, {Martini}, \& {Kaiser}}]{bird08}
{Bird}, J., {Martini}, P., \& {Kaiser}, C. 2008, \apj, 676, 147

\bibitem[{{B{\^i}rzan} {et~al.}(2017){B{\^i}rzan}, {Rafferty}, {Br{\"u}ggen},
  \& {Intema}}]{birzan17}
{B{\^i}rzan}, L., {Rafferty}, D.~A., {Br{\"u}ggen}, M., \& {Intema}, H.~T.
  2017, \mnras, 471, 1766

\bibitem[{{Bleem} {et~al.}(2015){Bleem}, {Stalder}, {de Haan}, {Aird}, {Allen},
  {Applegate}, {Ashby}, {Bautz}, {Bayliss}, {Benson}, {Bocquet}, {Brodwin},
  {Carlstrom}, {Chang}, {Chiu}, {Cho}, {Clocchiatti}, {Crawford}, {Crites},
  {Desai}, {Dietrich}, {Dobbs}, {Foley}, {Forman}, {George}, {Gladders},
  {Gonzalez}, {Halverson}, {Hennig}, {Hoekstra}, {Holder}, {Holzapfel},
  {Hrubes}, {Jones}, {Keisler}, {Knox}, {Lee}, {Leitch}, {Liu}, {Lueker},
  {Luong-Van}, {Mantz}, {Marrone}, {McDonald}, {McMahon}, {Meyer}, {Mocanu},
  {Mohr}, {Murray}, {Padin}, {Pryke}, {Reichardt}, {Rest}, {Ruel}, {Ruhl},
  {Saliwanchik}, {Saro}, {Sayre}, {Schaffer}, {Schrabback}, {Shirokoff},
  {Song}, {Spieler}, {Stanford}, {Staniszewski}, {Stark}, {Story}, {Stubbs},
  {Vanderlinde}, {Vieira}, {Vikhlinin}, {Williamson}, {Zahn}, \&
  {Zenteno}}]{bleem15}
{Bleem}, L.~E., {Stalder}, B., {de Haan}, T., {et~al.} 2015, \apjs, 216, 27

\bibitem[{{Bufanda} {et~al.}(2017){Bufanda}, {Hollowood}, {Jeltema}, {Rykoff},
  {Rozo}, {Martini}, {Abbott}, {Abdalla}, {Allam}, {Banerji},
  {Benoit-L{\'e}vy}, {Bertin}, {Brooks}, {Carnero Rosell}, {Carrasco Kind},
  {Carretero}, {Cunha}, {da Costa}, {Desai}, {Diehl}, {Dietrich}, {Evrard},
  {Fausti Neto}, {Flaugher}, {Frieman}, {Gerdes}, {Goldstein}, {Gruen},
  {Gruendl}, {Gutierrez}, {Honscheid}, {James}, {Kuehn}, {Kuropatkin}, {Lima},
  {Maia}, {Marshall}, {Melchior}, {Miquel}, {Mohr}, {Ogando}, {Plazas},
  {Romer}, {Rooney}, {Sanchez}, {Santiago}, {Scarpine}, {Sevilla-Noarbe},
  {Smith}, {Soares-Santos}, {Sobreira}, {Suchyta}, {Tarle}, {Thomas}, {Tucker},
  {Walker}, \& {DES Collaboration}}]{bufanda17}
{Bufanda}, E., {Hollowood}, D., {Jeltema}, T.~E., {et~al.} 2017, \mnras, 465,
  2531

\bibitem[{{Castignani} {et~al.}(2014){Castignani}, {Chiaberge}, {Celotti},
  {Norman}, \& {De Zotti}}]{castignani14}
{Castignani}, G., {Chiaberge}, M., {Celotti}, A., {Norman}, C., \& {De Zotti},
  G. 2014, \apj, 792, 114

\bibitem[{{Chabrier}(2003)}]{chabrier03}
{Chabrier}, G. 2003, \pasp, 115, 763

\bibitem[{{Chambers} {et~al.}(2016){Chambers}, {Magnier}, {Metcalfe},
  {Flewelling}, {Huber}, {Waters}, {Denneau}, {Draper}, {Farrow}, {Finkbeiner},
  {Holmberg}, {Koppenhoefer}, {Price}, {Saglia}, {Schlafly}, {Smartt},
  {Sweeney}, {Wainscoat}, {Burgett}, {Grav}, {Heasley}, {Hodapp}, {Jedicke},
  {Kaiser}, {Kudritzki}, {Luppino}, {Lupton}, {Monet}, {Morgan}, {Onaka},
  {Stubbs}, {Tonry}, {Banados}, {Bell}, {Bender}, {Bernard}, {Botticella},
  {Casertano}, {Chastel}, {Chen}, {Chen}, {Cole}, {Deacon}, {Frenk},
  {Fitzsimmons}, {Gezari}, {Goessl}, {Goggia}, {Goldman}, {Grebel}, {Hambly},
  {Hasinger}, {Heavens}, {Heckman}, {Henderson}, {Henning}, {Holman}, {Hopp},
  {Ip}, {Isani}, {Keyes}, {Koekemoer}, {Kotak}, {Long}, {Lucey}, {Liu},
  {Martin}, {McLean}, {Morganson}, {Murphy}, {Nieto-Santisteban}, {Norberg},
  {Peacock}, {Pier}, {Postman}, {Primak}, {Rae}, {Rest}, {Riess}, {Riffeser},
  {Rix}, {Roser}, {Schilbach}, {Schultz}, {Scolnic}, {Szalay}, {Seitz},
  {Shiao}, {Small}, {Smith}, {Soderblom}, {Taylor}, {Thakar}, {Thiel},
  {Thilker}, {Urata}, {Valenti}, {Walter}, {Watters}, {Werner}, {White},
  {Wood-Vasey}, \& {Wyse}}]{chambers16}
{Chambers}, K.~C., {Magnier}, E.~A., {Metcalfe}, N., {et~al.} 2016, ArXiv
  e-prints, arXiv:1612.05560

\bibitem[{{Chiaberge} {et~al.}(2015){Chiaberge}, {Gilli}, {Lotz}, \&
  {Norman}}]{chiaberge15}
{Chiaberge}, M., {Gilli}, R., {Lotz}, J.~M., \& {Norman}, C. 2015, \apj, 806,
  147

\bibitem[{{Chung} {et~al.}(2014){Chung}, {Kochanek}, {Assef}, {Brown}, {Stern},
  {Jannuzi}, {Gonzalez}, {Hickox}, \& {Moustakas}}]{chung14}
{Chung}, S.~M., {Kochanek}, C.~S., {Assef}, R., {et~al.} 2014, \apj, 790, 54

\bibitem[{{Condon} {et~al.}(1998){Condon}, {Cotton}, {Greisen}, {Yin},
  {Perley}, {Taylor}, \& {Broderick}}]{condon98}
{Condon}, J.~J., {Cotton}, W.~D., {Greisen}, E.~W., {et~al.} 1998, \aj, 115,
  1693

\bibitem[{{Conroy} {et~al.}(2009){Conroy}, {Gunn}, \& {White}}]{conroy09}
{Conroy}, C., {Gunn}, J.~E., \& {White}, M. 2009, \apj, 699, 486

\bibitem[{{Cutri} \& {et al.}(2013)}]{cutri13}
{Cutri}, R.~M., \& {et al.} 2013, VizieR Online Data Catalog, 2328, 0

\bibitem[{{Donoso} {et~al.}(2009){Donoso}, {Best}, \& {Kauffmann}}]{donoso09}
{Donoso}, E., {Best}, P.~N., \& {Kauffmann}, G. 2009, \mnras, 392, 617

\bibitem[{{Fabian}(2012)}]{fabian12}
{Fabian}, A.~C. 2012, \araa, 50, 455

\bibitem[{{Fanaroff} \& {Riley}(1974)}]{fanaroffriley}
{Fanaroff}, B.~L., \& {Riley}, J.~M. 1974, \mnras, 167, 31P

\bibitem[{{Galametz} {et~al.}(2009){Galametz}, {Stern}, {Eisenhardt},
  {Brodwin}, {Brown}, {Dey}, {Gonzalez}, {Jannuzi}, {Moustakas}, \&
  {Stanford}}]{galametz09}
{Galametz}, A., {Stern}, D., {Eisenhardt}, P.~R.~M., {et~al.} 2009, \apj, 694,
  1309

\bibitem[{{Gonzalez} {et~al.}(2019){Gonzalez}, {Gettings}, {Brodwin},
  {Eisenhardt}, {Stanford}, {Wylezalek}, {Decker}, {Marrone}, {Moravec},
  {O'Donnell}, {Stalder}, {Stern}, {Abdulla}, {Brown}, {Carlstrom}, {Chambers},
  {Hayden}, {Lin}, {Magnier}, {Masci}, {Mantz}, {McDonald}, {Mo}, {Perlmutter},
  {Wright}, \& {Zeimann}}]{gonzalez18}
{Gonzalez}, A.~H., {Gettings}, D.~P., {Brodwin}, M., {et~al.} 2019, \apjs, 240,
  33

\bibitem[{{Gralla} {et~al.}(2011){Gralla}, {Gladders}, {Yee}, \&
  {Barrientos}}]{gralla11}
{Gralla}, M.~B., {Gladders}, M.~D., {Yee}, H.~K.~C., \& {Barrientos}, L.~F.
  2011, \apj, 734, 103

\bibitem[{{Gralla} {et~al.}(2014){Gralla}, {Crichton}, {Marriage}, {Mo},
  {Aguirre}, {Addison}, {Asboth}, {Battaglia}, {Bock}, {Bond}, {Devlin},
  {D{\"u}nner}, {Hajian}, {Halpern}, {Hilton}, {Hincks}, {Hlozek},
  {Huffenberger}, {Hughes}, {Ivison}, {Kosowsky}, {Lin}, {Marsden},
  {Menanteau}, {Moodley}, {Morales}, {Niemack}, {Oliver}, {Page}, {Partridge},
  {Reese}, {Rojas}, {Sehgal}, {Sievers}, {Sif{\'o}n}, {Spergel}, {Staggs},
  {Switzer}, {Viero}, {Wollack}, \& {Zemcov}}]{gralla14}
{Gralla}, M.~B., {Crichton}, D., {Marriage}, T.~A., {et~al.} 2014, \mnras, 445,
  460

\bibitem[{{Gupta} {et~al.}(2017){Gupta}, {Saro}, {Mohr}, {Benson}, {Bocquet},
  {Capasso}, {Carlstrom}, {Chiu}, {Crawford}, {de Haan}, {Dietrich},
  {Gangkofner}, {Holzapfel}, {McDonald}, {Rapetti}, \& {Reichardt}}]{gupta17}
{Gupta}, N., {Saro}, A., {Mohr}, J.~J., {et~al.} 2017, \mnras, 467, 3737

\bibitem[{{Hambly} {et~al.}(2001){Hambly}, {MacGillivray}, {Read}, {Tritton},
  {Thomson}, {Kelly}, {Morgan}, {Smith}, {Driver}, {Williamson}, {Parker},
  {Hawkins}, {Williams}, \& {Lawrence}}]{hambly01}
{Hambly}, N.~C., {MacGillivray}, H.~T., {Read}, M.~A., {et~al.} 2001, \mnras,
  326, 1279

\bibitem[{{Hardcastle} {et~al.}(2019){Hardcastle}, {Williams}, {Best},
  {Croston}, {Duncan}, {R{\"o}ttgering}, {Sabater}, {Shimwell}, {Tasse},
  {Callingham}, {Cochrane}, {de Gasperin}, {G{\"u}rkan}, {Jarvis}, {Mahatma},
  {Miley}, {Mingo}, {Mooney}, {Morabito}, {O'Sullivan}, {Prandoni},
  {Shulevski}, \& {Smith}}]{hardcastle19}
{Hardcastle}, M.~J., {Williams}, W.~L., {Best}, P.~N., {et~al.} 2019, \aap,
  622, A12

\bibitem[{{Hasselfield} {et~al.}(2013){Hasselfield}, {Hilton}, {Marriage},
  {Addison}, {Barrientos}, {Battaglia}, {Battistelli}, {Bond}, {Crichton},
  {Das}, {Devlin}, {Dicker}, {Dunkley}, {D{\"u}nner}, {Fowler}, {Gralla},
  {Hajian}, {Halpern}, {Hincks}, {Hlozek}, {Hughes}, {Infante}, {Irwin},
  {Kosowsky}, {Marsden}, {Menanteau}, {Moodley}, {Niemack}, {Nolta}, {Page},
  {Partridge}, {Reese}, {Schmitt}, {Sehgal}, {Sherwin}, {Sievers}, {Sif{\'o}n},
  {Spergel}, {Staggs}, {Swetz}, {Switzer}, {Thornton}, {Trac}, \&
  {Wollack}}]{hasselfield13}
{Hasselfield}, M., {Hilton}, M., {Marriage}, T.~A., {et~al.} 2013, \jcap, 7,
  008

\bibitem[{{Hatch} {et~al.}(2014){Hatch}, {Wylezalek}, {Kurk}, {Stern}, {De
  Breuck}, {Jarvis}, {Galametz}, {Gonzalez}, {Hartley}, {Mortlock}, {Seymour},
  \& {Stevens}}]{hatch14}
{Hatch}, N.~A., {Wylezalek}, D., {Kurk}, J.~D., {et~al.} 2014, \mnras, 445, 280

\bibitem[{{Hinshaw} {et~al.}(2013){Hinshaw}, {Larson}, {Komatsu}, {Spergel},
  {Bennett}, {Dunkley}, {Nolta}, {Halpern}, {Hill}, {Odegard}, {Page}, {Smith},
  {Weiland}, {Gold}, {Jarosik}, {Kogut}, {Limon}, {Meyer}, {Tucker}, {Wollack},
  \& {Wright}}]{hinshaw13}
{Hinshaw}, G., {Larson}, D., {Komatsu}, E., {et~al.} 2013, \apjs, 208, 19

\bibitem[{{Hlavacek-Larrondo} {et~al.}(2013){Hlavacek-Larrondo}, {Allen},
  {Taylor}, {Fabian}, {Canning}, {Werner}, {Sanders}, {Grimes}, {Ehlert}, \&
  {von der Linden}}]{hlavacek-larrondo13}
{Hlavacek-Larrondo}, J., {Allen}, S.~W., {Taylor}, G.~B., {et~al.} 2013, \apj,
  777, 163

\bibitem[{{Intema} {et~al.}(2017){Intema}, {Jagannathan}, {Mooley}, \&
  {Frail}}]{intema17}
{Intema}, H.~T., {Jagannathan}, P., {Mooley}, K.~P., \& {Frail}, D.~A. 2017,
  \aap, 598, A78

\bibitem[{{Izquierdo-Villalba} {et~al.}(2018){Izquierdo-Villalba}, {Orsi},
  {Bonoli}, {Lacey}, {Baugh}, \& {Griffin}}]{izquierdo-villalba18}
{Izquierdo-Villalba}, D., {Orsi}, {\'A}.~A., {Bonoli}, S., {et~al.} 2018,
  \mnras, 480, 1340

\bibitem[{{Joshi} {et~al.}(2017){Joshi}, {Wadsley}, \& {Parker}}]{joshi17}
{Joshi}, G.~D., {Wadsley}, J., \& {Parker}, L.~C. 2017, \mnras, 468, 4625

\bibitem[{{Kale} {et~al.}(2015){Kale}, {Venturi}, {Cassano}, {Giacintucci},
  {Bardelli}, {Dallacasa}, \& {Zucca}}]{kale15}
{Kale}, R., {Venturi}, T., {Cassano}, R., {et~al.} 2015, \aap, 581, A23

\bibitem[{{Kellermann} {et~al.}(2016){Kellermann}, {Condon}, {Kimball},
  {Perley}, \& {Ivezi{\'c}}}]{kellerman16}
{Kellermann}, K.~I., {Condon}, J.~J., {Kimball}, A.~E., {Perley}, R.~A., \&
  {Ivezi{\'c}}, {\v Z}. 2016, \apj, 831, 168

\bibitem[{{Lin} \& {Mohr}(2007)}]{lin07}
{Lin}, Y.-T., \& {Mohr}, J.~J. 2007, \apjs, 170, 71

\bibitem[{{Lin} {et~al.}(2009){Lin}, {Partridge}, {Pober}, {Bouchefry},
  {Burke}, {Klein}, {Coish}, \& {Huffenberger}}]{lin09}
{Lin}, Y.-T., {Partridge}, B., {Pober}, J.~C., {et~al.} 2009, \apj, 694, 992

\bibitem[{{Lin} {et~al.}(2017){Lin}, {Hsieh}, {Lin}, {Oguri}, {Chen}, {Tanaka},
  {Chiu}, {Huang}, {Kodama}, {Leauthaud}, {More}, {Nishizawa}, {Bundy}, {Lin},
  \& {Miyazaki}}]{lin17}
{Lin}, Y.-T., {Hsieh}, B.-C., {Lin}, S.-C., {et~al.} 2017, \apj, 851, 139

\bibitem[{{Mancone} \& {Gonzalez}(2012)}]{mancone12}
{Mancone}, C.~L., \& {Gonzalez}, A.~H. 2012, \pasp, 124, 606

\bibitem[{{Mancone} {et~al.}(2010){Mancone}, {Gonzalez}, {Brodwin}, {Stanford},
  {Eisenhardt}, {Stern}, \& {Jones}}]{mancone10}
{Mancone}, C.~L., {Gonzalez}, A.~H., {Brodwin}, M., {et~al.} 2010, \apj, 720,
  284

\bibitem[{{Martini} {et~al.}(2013){Martini}, {Miller}, {Brodwin}, {Stanford},
  {Gonzalez}, {Bautz}, {Hickox}, {Stern}, {Eisenhardt}, {Galametz}, {Norman},
  {Jannuzi}, {Dey}, {Murray}, {Jones}, \& {Brown}}]{martini13}
{Martini}, P., {Miller}, E.~D., {Brodwin}, M., {et~al.} 2013, \apj, 768, 1

\bibitem[{{McDonald} {et~al.}(2015){McDonald}, {McNamara}, {van Weeren},
  {Applegate}, {Bayliss}, {Bautz}, {Benson}, {Carlstrom}, {Bleem}, {Chatzikos},
  {Edge}, {Fabian}, {Garmire}, {Hlavacek-Larrondo}, {Jones-Forman}, {Mantz},
  {Miller}, {Stalder}, {Veilleux}, \& {ZuHone}}]{mcdonald15}
{McDonald}, M., {McNamara}, B.~R., {van Weeren}, R.~J., {et~al.} 2015, \apj,
  811, 111

\bibitem[{{Miley} \& {De Breuck}(2008)}]{miley08}
{Miley}, G., \& {De Breuck}, C. 2008, \aapr, 15, 67

\bibitem[{{Mittal} {et~al.}(2009){Mittal}, {Hudson}, {Reiprich}, \&
  {Clarke}}]{mittal09}
{Mittal}, R., {Hudson}, D.~S., {Reiprich}, T.~H., \& {Clarke}, T. 2009, \aap,
  501, 835

\bibitem[{{Mo} {et~al.}(2018){Mo}, {Gonzalez}, {Stern}, {Brodwin}, {Decker},
  {Eisenhardt}, {Moravec}, {Stanford}, \& {Wylezalek}}]{mo18}
{Mo}, W., {Gonzalez}, A., {Stern}, D., {et~al.} 2018, \apj, 869, 131

\bibitem[{{Moravec} {et~al.}(2019){Moravec}, {Gonzalez}, {Stern}, {Brodwin},
  {Clarke}, {Decker}, {Eisenhardt}, {Mo}, {O'Donnell}, {Pope}, {Stanford}, \&
  {Wylezalek}}]{moravec19}
{Moravec}, E., {Gonzalez}, A.~H., {Stern}, D., {et~al.} 2019, \apj, 871, 186

\bibitem[{{Moravec} {et~al.}(2020{\natexlab{a}}){Moravec}, {Gonzalez},
  {Dicker}, {Alberts}, {Brodwin}, {Clarke}, {Connor}, {Decker}, {Devlin},
  {Eisenhardt}, {Mason}, {Mo}, {Mroczkowski}, {Pope}, {Romero}, {Sarazin},
  {Sievers}, {Stanford}, {Stern}, {Wylezalek}, \& {Zago}}]{moravec20b}
{Moravec}, E., {Gonzalez}, A.~H., {Dicker}, S., {et~al.} 2020{\natexlab{a}},
  \apj, 898, 145

\bibitem[{{Moravec} {et~al.}(2020{\natexlab{b}}){Moravec}, {Gonzalez}, {Stern},
  {Clarke}, {Brodwin}, {Decker}, {Eisenhardt}, {Mo}, {Pope}, {Stanford}, \&
  {Wylezalek}}]{moravec20}
{Moravec}, E., {Gonzalez}, A.~H., {Stern}, D., {et~al.} 2020{\natexlab{b}},
  \apj, 888, 74

\bibitem[{{Muzzin} {et~al.}(2013){Muzzin}, {Wilson}, {Demarco}, {Lidman},
  {Nantais}, {Hoekstra}, {Yee}, \& {Rettura}}]{muzzin13}
{Muzzin}, A., {Wilson}, G., {Demarco}, R., {et~al.} 2013, \apj, 767, 39

\bibitem[{{Noirot} {et~al.}(2016){Noirot}, {Vernet}, {De Breuck}, {Wylezalek},
  {Galametz}, {Stern}, {Mei}, {Brodwin}, {Cooke}, {Gonzalez}, {Hatch},
  {Rettura}, \& {Stanford}}]{noirot16}
{Noirot}, G., {Vernet}, J., {De Breuck}, C., {et~al.} 2016, \apj, 830, 90

\bibitem[{{Noirot} {et~al.}(2018){Noirot}, {Stern}, {Mei}, {Wylezalek},
  {Cooke}, {De Breuck}, {Galametz}, {Hatch}, {Vernet}, {Brodwin}, {Eisenhardt},
  {Gonzalez}, {Jarvis}, {Rettura}, {Seymour}, \& {Stanford}}]{noirot18}
{Noirot}, G., {Stern}, D., {Mei}, S., {et~al.} 2018, \apj, 859, 38

\bibitem[{{Orsi} {et~al.}(2016){Orsi}, {Fanidakis}, {Lacey}, \&
  {Baugh}}]{orsi16}
{Orsi}, {\'A}.~A., {Fanidakis}, N., {Lacey}, C.~G., \& {Baugh}, C.~M. 2016,
  \mnras, 456, 3827

\bibitem[{{Owen} {et~al.}(1999){Owen}, {Ledlow}, {Keel}, \&
  {Morrison}}]{owen99}
{Owen}, F.~N., {Ledlow}, M.~J., {Keel}, W.~C., \& {Morrison}, G.~E. 1999, \aj,
  118, 633

\bibitem[{{Papovich}(2008)}]{papovich08}
{Papovich}, C. 2008, \apj, 676, 206

\bibitem[{{Paterno-Mahler} {et~al.}(2017){Paterno-Mahler}, {Blanton},
  {Brodwin}, {Ashby}, {Golden-Marx}, {Decker}, {Wing}, \&
  {Anand}}]{paterno-mahler17}
{Paterno-Mahler}, R., {Blanton}, E.~L., {Brodwin}, M., {et~al.} 2017, \apj,
  844, 78

\bibitem[{{Richards} {et~al.}(2015){Richards}, {Myers}, {Peters}, {Krawczyk},
  {Chase}, {Ross}, {Fan}, {Jiang}, {Lacy}, {McGreer}, {Trump}, \&
  {Riegel}}]{richards15}
{Richards}, G.~T., {Myers}, A.~D., {Peters}, C.~M., {et~al.} 2015, \apjs, 219,
  39

\bibitem[{{Rigby} {et~al.}(2011){Rigby}, {Best}, {Brookes}, {Peacock},
  {Dunlop}, {R{\"o}ttgering}, {Wall}, \& {Ker}}]{rigby11}
{Rigby}, E.~E., {Best}, P.~N., {Brookes}, M.~H., {et~al.} 2011, \mnras, 416,
  1900

\bibitem[{{Rozo} {et~al.}(2009){Rozo}, {Rykoff}, {Koester}, {McKay}, {Hao},
  {Evrard}, {Wechsler}, {Hansen}, {Sheldon}, {Johnston}, {Becker}, {Annis},
  {Bleem}, \& {Scranton}}]{rozo09}
{Rozo}, E., {Rykoff}, E.~S., {Koester}, B.~P., {et~al.} 2009, \apj, 703, 601

\bibitem[{{Rykoff} {et~al.}(2012){Rykoff}, {Koester}, {Rozo}, {Annis},
  {Evrard}, {Hansen}, {Hao}, {Johnston}, {McKay}, \& {Wechsler}}]{rykoff12}
{Rykoff}, E.~S., {Koester}, B.~P., {Rozo}, E., {et~al.} 2012, \apj, 746, 178

\bibitem[{{Rykoff} {et~al.}(2014){Rykoff}, {Rozo}, {Busha}, {Cunha},
  {Finoguenov}, {Evrard}, {Hao}, {Koester}, {Leauthaud}, {Nord}, {Pierre},
  {Reddick}, {Sadibekova}, {Sheldon}, \& {Wechsler}}]{rykoff14}
{Rykoff}, E.~S., {Rozo}, E., {Busha}, M.~T., {et~al.} 2014, \apj, 785, 104

\bibitem[{{Seymour} {et~al.}(2007){Seymour}, {Stern}, {De Breuck}, {Vernet},
  {Rettura}, {Dickinson}, {Dey}, {Eisenhardt}, {Fosbury}, {Lacy}, {McCarthy},
  {Miley}, {Rocca-Volmerange}, {R{\"o}ttgering}, {Stanford}, {Teplitz}, {van
  Breugel}, \& {Zirm}}]{seymour07}
{Seymour}, N., {Stern}, D., {De Breuck}, C., {et~al.} 2007, \apjs, 171, 353

\bibitem[{{Smol{\v c}i{\'c}} {et~al.}(2009){Smol{\v c}i{\'c}}, {Zamorani},
  {Schinnerer}, {Bardelli}, {Bondi}, {B{\^i}rzan}, {Carilli}, {Ciliegi},
  {Elvis}, {Impey}, {Koekemoer}, {Merloni}, {Paglione}, {Salvato}, {Scodeggio},
  {Scoville}, \& {Trump}}]{smolcic09}
{Smol{\v c}i{\'c}}, V., {Zamorani}, G., {Schinnerer}, E., {et~al.} 2009, \apj,
  696, 24

\bibitem[{{Sobral} {et~al.}(2015){Sobral}, {Stroe}, {Dawson}, {Wittman}, {Jee},
  {R{\"o}ttgering}, {van Weeren}, \& {Br{\"u}ggen}}]{sobral15}
{Sobral}, D., {Stroe}, A., {Dawson}, W.~A., {et~al.} 2015, \mnras, 450, 630

\bibitem[{{Sommer} {et~al.}(2011){Sommer}, {Basu}, {Pacaud}, {Bertoldi}, \&
  {Andernach}}]{sommer11}
{Sommer}, M.~W., {Basu}, K., {Pacaud}, F., {Bertoldi}, F., \& {Andernach}, H.
  2011, \aap, 529, A124

\bibitem[{{Stern} {et~al.}(2012){Stern}, {Assef}, {Benford}, {Blain}, {Cutri},
  {Dey}, {Eisenhardt}, {Griffith}, {Jarrett}, {Lake}, {Masci}, {Petty},
  {Stanford}, {Tsai}, {Wright}, {Yan}, {Harrison}, \& {Madsen}}]{stern12}
{Stern}, D., {Assef}, R.~J., {Benford}, D.~J., {et~al.} 2012, \apj, 753, 30

\bibitem[{{Stott} {et~al.}(2012){Stott}, {Hickox}, {Edge}, {Collins}, {Hilton},
  {Harrison}, {Romer}, {Rooney}, {Kay}, {Miller}, {Sahl{\'e}n}, {Lloyd-Davies},
  {Mehrtens}, {Hoyle}, {Liddle}, {Viana}, {McCarthy}, {Schaye}, \&
  {Booth}}]{stott12}
{Stott}, J.~P., {Hickox}, R.~C., {Edge}, A.~C., {et~al.} 2012, \mnras, 422,
  2213

\bibitem[{{Wing} \& {Blanton}(2011)}]{wing11}
{Wing}, J.~D., \& {Blanton}, E.~L. 2011, \aj, 141, 88

\bibitem[{{Wright} {et~al.}(2010){Wright}, {Eisenhardt}, {Mainzer}, {Ressler},
  {Cutri}, {Jarrett}, {Kirkpatrick}, {Padgett}, {McMillan}, {Skrutskie},
  {Stanford}, {Cohen}, {Walker}, {Mather}, {Leisawitz}, {Gautier}, {McLean},
  {Benford}, {Lonsdale}, {Blain}, {Mendez}, {Irace}, {Duval}, {Liu}, {Royer},
  {Heinrichsen}, {Howard}, {Shannon}, {Kendall}, {Walsh}, {Larsen}, {Cardon},
  {Schick}, {Schwalm}, {Abid}, {Fabinsky}, {Naes}, \& {Tsai}}]{wright10}
{Wright}, E.~L., {Eisenhardt}, P.~R.~M., {Mainzer}, A.~K., {et~al.} 2010, \aj,
  140, 1868

\bibitem[{{Wylezalek} {et~al.}(2013){Wylezalek}, {Galametz}, {Stern}, {Vernet},
  {De Breuck}, {Seymour}, {Brodwin}, {Eisenhardt}, {Gonzalez}, {Hatch},
  {Jarvis}, {Rettura}, {Stanford}, \& {Stevens}}]{wylezalek13}
{Wylezalek}, D., {Galametz}, A., {Stern}, D., {et~al.} 2013, \apj, 769, 79

\bibitem[{{Wylezalek} {et~al.}(2014){Wylezalek}, {Vernet}, {De Breuck},
  {Stern}, {Brodwin}, {Galametz}, {Gonzalez}, {Jarvis}, {Hatch}, {Seymour}, \&
  {Stanford}}]{wylezalek14}
{Wylezalek}, D., {Vernet}, J., {De Breuck}, C., {et~al.} 2014, \apj, 786, 17

\bibitem[{{Yuan} {et~al.}(2016){Yuan}, {Han}, \& {Wen}}]{yuan16}
{Yuan}, Z.~S., {Han}, J.~L., \& {Wen}, Z.~L. 2016, \mnras, 460, 3669

\end{thebibliography}
\end{document}